\documentclass[letterpaper, 10 pt, journal, twocolumn, twoside]{support/IEEEtran}
\usepackage[fleqn]{amsmath}
\usepackage{times}
\usepackage[pdftex]{graphicx}
\usepackage{subfigure}
\usepackage{amsmath,amssymb,amsopn,amstext,amsfonts}
\usepackage{cancel}
\usepackage[noadjust]{cite}
\usepackage{soul}
\usepackage{caption}
\usepackage{array}

\usepackage{mathrsfs}
\usepackage{amssymb}
\usepackage{balance}
\usepackage{color}
\usepackage{enumerate}
\usepackage{indentfirst}
\usepackage{mathtools}
\usepackage{algorithm}
\usepackage{dsfont}
\usepackage{bm}
\usepackage{diagbox}
\usepackage{float}
\usepackage{epstopdf}
\usepackage{epsfig}
\usepackage{graphics}
\usepackage{graphicx}
\usepackage{setspace}
\usepackage{url}
\usepackage{multirow}
\usepackage{multicol}
\usepackage{tikz}
\usepackage{subeqnarray}
\usepackage{cases}
\usepackage{booktabs}
\usepackage[linkcolor=black,citecolor=black,urlcolor=black,colorlinks=true]{hyperref}
\usepackage{algorithm}
\usepackage[noend]{algpseudocode}
\newtheorem{myTheo}{Theorem}
\newtheorem{myDef}{Definition} 
\newtheorem{lemma}{Lemma} 
\newtheorem{myCollo}{Corollary}
\newtheorem{remark}{Remark}

\newtheorem{assumption}{Assumption}

\soulregister\cite7
\soulregister\citep7
\soulregister\citet7
\soulregister\ref7
\soulregister\it7
\soulregister\pageref7

\normalbaselineskip=\baselineskip
\renewcommand{\baselinestretch}{1.0}
\makeatletter
\newcommand*\bigcdot{\mathpalette\bigcdot@{.5}}
\newcommand*\bigcdot@[2]{\mathbin{\vcenter{\hbox{\scalebox{#2}{$\m@th#1\bullet$}}}}}
\makeatother

\graphicspath{{figures/}}
\DeclareGraphicsExtensions{.pdf,.png,.jpg,.eps}
\IEEEoverridecommandlockouts

\title{\LARGE \bf Optimal Resource Allocation between Two Nonfully Cooperative Wireless Networks under Malicious Attacks: A Gestalt Game Perspective}

\author{
  \vskip 1em
  {
  Yukang Cui, 
  Xinru Yang,
  Tingwen Huang, 
  and Xin Gong
  }

  \thanks{
    This work was partially supported by the National Natural Science Foundation of China under Grant 61903258, Guangdong Basic and Applied Basic Research Foundation 2022A1515010234 and the Project of Department of Education of Guangdong Province 2022KTSCX105.(\emph{Corresponding author: Xin
    Gong.}) 
    
Y. Cui and X. Yang are with the College of Mechatronics and Control Engineering, Shenzhen University, Shenzhen, 518060, China (e-mail: {\tt\small cuiyukang,szuwtn@gmail.com}).

T. Huang is with Texas A\&M University at Qatar, Doha, 23874, Qatar (e-mail: {\tt\small tingwen.huang@qatar.tamu.edu}).

X. Gong is with the Department of Mechanical Engineering, The University of Hong Kong, Pokfulam Road, Hong Kong (e-mail: {\tt\small gongxin@connect.hku.hk}).

  }
}

\begin{document}
\maketitle
\begin{abstract}
In this paper, the problem of seeking optimal distributed resource allocation (DRA) policies on cellular networks in the presence of an unknown malicious adding-edge attacker is investigated. This problem is described as the games of games (GoG) model. Specifically, two subnetwork policymakers constitute a Nash game, while the confrontation between each subnetwork policymaker and the attacker is captured by a Stackelberg game. First, we show that the communication resource allocation of cellular networks based on the Foschini-Miljanic (FM) algorithm can be transformed into a \emph{geometric program} and be efficiently solved via convex optimization. 
Second, the upper limit of attack magnitude that can be tolerated by the network is calculated by the corresponding theory, and it is proved that the above geometric programming (GP) framework is solvable within the attack bound, that is, there exists a Gestalt Nash equilibrium (GNE) in our GoG.
Third, a heuristic algorithm that iteratively uses GP is proposed to identify the optimal policy profiles of both subnetworks, for which asymptotic convergence is also confirmed. Fourth, a greedy heuristic adding-edge strategy is developed for the attacker to determine the set of the most vulnerable edges. Finally, simulation examples illustrate that the proposed theoretical results are robust and can achieve the GNE. It is verified that the transmission gains and interference gains of all channels are well tuned within a limited budget, despite the existence of malicious attacks.

\end{abstract}

\section{Introduction}\label{section1}
In wireless cellular networks, stable and efficient data transmission requires the reasonable and robust resource allocation to achieve quality and bandwidth efficiency, especially now that wireless networks support a rapidly expanding range of applications with different quality of service (QoS) constraints. This resource allocation problem can usually be formulated as an optimization problem 
of network-wide cost functions.
The related distributed resource allocation (DRA) \cite{Li2013Communication} problem has become a popular research topic in various research directions, such as energy-efficient resource allocation \cite{2007Energy}, network capacity optimization \cite{2007Adaptation}, and network sum-rate maximization \cite{6848847}. It is a promising approach that achieves a good balance between excellent QoS and lower energy costs for cellular networks.
As a representative optimal DRA scheme, the well-known Foschini-Miljanic (FM) algorithm in \cite{1993A} is of fundamental importance. This distributed algorithm effectively controls interference in a new type of multiple access technology developed for 5G, that is, sparse code multiple access, and maximizes the degree of channel multiplexing while achieving the signal-to-interference-plus-noise-ratio (SINR) expected by cellular users.

In this paper, a method to solve the resource allocation problem for realizing interference containment of user equipment in a wireless cellular network consisting of two non-fully cooperative subnetworks is investigated on the basis of the FM algorithm. Once a set of power settings exists that satisfies the overall QoS requirements of the network, the algorithm rapidly converges to that power at an exponential rate. Moreover, an unknown but bounded malicious adding-edge attacker that poses a threat to network security is also considered.
Each policymaker can allocate communication resources to the nodes in the subnetwork under his control when attacked. In this work, the communication resources are used to tune two parameters in the FM algorithm to achieve dynamic stability: 
1) The transmission gain of the channel.
2) The interference gain among the channels.
It is assumed that the adjustment of these two gains has its corresponding cost function.
Based on the above settings, we study the problem of decentralized communication resource allocation with cost-optimal advantages to curb the impact of interference according to the SINR prerequired by cellular users.

To this end, a games of games (GoG) framework, also known as \emph{Gestalt Games} \cite{chen2018security}, is developed. It is a resilient DRA design problem that takes into account both subnet-subnet interactions and subnet-attacker confrontations. The former is captured by a \emph{Nash game} \cite{2013Game} to describe the lack of coordination between the two subnetworks; the latter is captured by a \emph{Stackelberg game} \cite{maharjan2013dependable} to describe how each policymaker allocates communication resources by anticipating a set of critical edges added by the malicious attacker. It is found that there may be a \emph{Gestalt Nash equilibrium} (GNE) \cite{chen2018security} in the above GoG, which is composed of the optimal resource allocation policy of each subnetwork policymaker and the corresponding adversarial adding-edge strategy.

\subsection{Related Works}

Numerous works \cite{2002CDMA, 2004Opportunistic, tan2007exploiting, xu2016distributed, 8870437, ramezani2017joint, huo2019stackelberg, 7222480, 7317725, chen2021hybrid, zappone2015energy, ogura2019resource, 7328326, 7862919, 8626185} have investigated how to allocate communication resources for interference control to support the SINR required by cellular users.
This problem is typically formulated as optimization problems with cost functions for a given network topology.
Hidden convexity in seemingly nonconvex problems is investigated in \cite{tan2007exploiting}, such that the resource allocation in wireless networks is to be flexibly and stably adjusted in a distributed manner.
The conversion of the nonconvex form of iterative resource allocation and power control optimization problems into an equivalent optimization problem in the subtraction form is investigated in \cite{7222480}, however, it cannot be solved as efficiently as convex optimization.
Many studies have employed various optimization techniques, such as integer programming \cite{huo2019stackelberg}, fractional programming \cite{7222480}, nonlinear fractional programming \cite{7317725}, mixed integer nonlinear programming \cite{chen2021hybrid} and geometric programming (GP) \cite{tan2007exploiting}, \cite{ogura2019resource}, to solve the optimal allocation scheme of distributed limited communication resources.
Zappone \emph{et al.} developed various optimization algorithms to maximize energy efficiency in wireless networks in different scenarios through fractional programming \cite{zappone2015energy}, game theory \cite{7328326}, and a combination of fractional programming and sequential optimization \cite{7862919}.
An efficient channel assignment algorithm is proposed \cite{8626185}, in which the Markov decision process outperforms random assignment. This work is related to \cite{2002CDMA}, \cite{tan2007exploiting}, and \cite{ogura2019resource}, where the authors are committed to achieving interference control with a lower budget to reach the desired SINR.
Compared with the work of \cite{2002CDMA}, \cite{tan2007exploiting}, and \cite{ogura2019resource}, two main improvements are as follows:
\begin{enumerate}
  \item Our work further considers the compromised network topology distorted by an unknown but bounded attacker.
  \item The conflict between two subnetworks is taken into account, where policymakers make decisions in a nonfully cooperative manner.
\end{enumerate}
Therefore, the problem of communication resource allocation is more realistic and challenging in our environment.


However, the above DRA optimization problems are usually nonconvex, which makes it difficult to obtain a globally optimal solution. Therefore, in Section \ref{section3}, geometric programming (GP) is introduced to optimize the polynomial function objectives and constraint sets with structured uncertainties. Based on these results, the framework of GP is generalized to the problem of globally optimal resource allocation in the presence of malicious attacks. It is proven that the robust resource allocation to the objective function under various constraints can be effectively regulated.

GP \cite{Stephen2007A} is a nonlinear optimization scheme whose objective function and constraints consist of generalized  \emph{posynomials}.
It is widely used in various fields, such as the optimization design in the chemical industry \cite{wall1986or}, power control \cite{4275017}, and resource allocation \cite{ogura2019geometric, ogura2019resource}. Since GP contains and optimizes only positive parameters, it plays a crucial role in the positive system \cite{cui2021positivity}. GP-based algorithms generally are robust and time-efficient in solving.
As shown in \cite{ogura2019geometric}, the optimization framework for parameter tuning problems constrained by $H_2$ norm, $H_{\infty}$ norm, Hankel norm or Schattern $p$-norm can be established and solved efficiently via GP. 
For the uncertainty of the log-quantized feedback errors in the FM algorithm, the stabilization problem for bounded-input bounded-output systems is investigated in \cite{8931743}. The refinement of the quantization level enables the cellular network to obtain a better QoS.
In \cite{colombino2015convex}, the robust stabilization of the FM algorithm when uncertainty exists in the disturbance is given by the linear matrix inequality (LMI). Compared with LMI, GP allows greater scalability since it can solve more complex and larger-scale networks with higher accuracy.
Therefore, the robust stabilization and resource allocation problems of the FM algorithm with structured uncertainties are further investigated through GP in \cite{ogura2019resource}.
In this work, we propose a convex optimization framework, specifically GP, for the robust stabilization problem under structured uncertainties of the discrete-time FM algorithm. The GP formulation is extended to an iterative algorithm to address the resilient stabilization problem of cellular network QoS under the threat of adding-edge attacks, that is, 
to determine the precise GNE of the two subnetwork policymakers.
The attacker can be regarded as the worst-case structured uncertainty in this GP framework with norm constraints.
Compared with the Stackelberg game framework in \cite{2002CDMA} that allocates a single resource, a key advantage of the GP scheme is that the resource allocation for both gains can be optimized simultaneously as shown in Lemma \ref{Y-the-1}.

\subsection{Main Contributions}
\begin{enumerate}
  \item 
  Against the backdrop of adding-edge attacks on two communication subnetworks, a framework for solving the GoG is constructed. It is demonstrated that the robust stabilization problem of the discrete-time FM algorithm under norm-bounded structured uncertainties can be transformed into a standard GP problem (see Lemma \ref{Y-the-1}). Since the existence of the attacker can be viewed as the worst case of structured uncertainties, the above GP framework can be utilized as a basis to solve the problem of interference suppression under the threat of malicious attacks.
  \item \textbf{From the defender side,} the lack of coordination between two policymakers caused by decentralized decision making is considered. Assume that they make decisions in a non-fully cooperative manner. The tolerable maximum magnitude of the attacks is computed via the GP scheme, and the GNE is proven to exist under this bound. Then, a GP-based round-robin heuristic algorithm that asymptotically converges to the GNE over multiple iterations is proposed.
  \item \textbf{From the attacker side,}  we develop a greedy heuristic strategy to increase the topological connectivity by adding a set of key edges through iterative selection; its attack mode and capability are taken into account.
\end{enumerate}

\noindent\textbf{Notations:}
We define $\mathbb{R}$, $\mathbb{R}_{\geq 0}$, $\mathbb{R}_{> 0} $,
$ \mathbb{Z}_{> 0} $ as the sets of real numbers, nonnegative real numbers, positive numbers, and positive integers, respectively.
The index set of sequential integers is denoted by $\textbf{I}[m,n]=\{m,m+1,\dotsc,n\}$, with $m,n\in\mathbb{Z}_{\geq 0}$ satisfying $m < n$.
The diagonal matrix of scalar diagonals $ a_i $  is denoted by ${\rm diag}(a_1, a_2, \dotsc, a_n)$. Similarly, the block diagonal matrix whose diagonals are composed of block diagonal matrices $A_i$, $i\in \textbf{I}[1,n]$, is denoted by $ {\rm blkdiag}(A_1, A_2, \dotsc, A_n) $.
For a vector $ \kappa = (k_1, k_2, \dotsc, k_n)^\mathrm{T} $, we use the identifier $ D_\kappa = {\rm diag}(k_1, k_2, \dotsc, k_n) $. $ I_n $ (or $ O_n $) represents the unit (or zero) matrix of order $ n $.
A column vector of size $ n $ with all entries equal to $ 1 $ is denoted by $ \boldsymbol{1}_n$. Define the componentwise exponent and logarithm of a vector $ \textbf{x} = [x_1 \dots x_n]^\mathrm{T} \in \mathbb{R}^n_{> 0}$ as $\exp[\textbf{x}] = [\exp (x_1) \dots \exp (x_n)]^\mathrm{T}$ and $\log[\textbf{x}] = [\log (x_1) \dots \log (x_n)]^\mathrm{T}$, respectively.

A matrix $ A\in \mathbb{R}^{n\times n}$ is called \emph{Metzler} if its off-diagonal entries are all nonnegative, that is,
$ [A]_{i j} \geq 0 $, $\forall i \neq j $. The notation $\lambda_i(A)$, $i\in \textbf{I}[1,n]$, denotes eigenvalues of $ A $, which can be sorted as $\lambda_1(A)\leq\lambda_2(A) \leq\ldots \leq\lambda_n(A)$.
Moreover, $\lambda_{\max}(A)$ represents the maximum real eigenvalue of $ A $. If all $\lambda_i(A)$ have negative real parts, $ A $ is called \emph{Hurwitz}.
If all $ \lambda_i(A) $ are located within the closed unit circle, $ A $ is called $ \emph{Schur stable} $.
Denote the Euclidean norm of a vector $ \nu $ as $ \|\nu \| $.
A \emph{nonnegative} (or \emph{positive}) matrix $ A $ is denoted by $ A \succeq 0\, $ (or $ A \succ 0 $), implying that all entries of $ A $ are nonnegative (or positive).
According to the \emph{Perron-Frobenius} theorem \cite{horn2012matrix}, there exists $ \lambda_{\max}(A) \geq \max_{i \in \textbf{I}[1,n]}{{\rm Re}(\lambda_i(A))} $ of a Metzler matrix $ A $, where $ {\rm Re}(\cdot) $ denotes the calculation of the real part. 
Moreover, the unit eigenvector $\mu$ corresponding to the above $\lambda_{\max}(A)$ is called the \emph{right Perron vector}, which follows $\mu > 0$ and $A \mu = \lambda_{\max}(A) \mu$.
For a given matrix $ A \in \mathbb{R}^{m\times n} $, its 1-norm, 2-norm and $ L_1 $-norm are denoted by $\|A\|_{1}=\max_{j\in \textbf{I}[1,n]}(\sum_{i=1}^{m}|[A]_{ij}|)$, $\|A\|_{2}=\sqrt{\lambda_{\max}(A^{\mathrm{T}}A)}$ and $\|A\|_{L_1}=\sum_{i=1}^{m}\sum_{j=1}^{n}|[A]_{ij}|$, respectively.
If a matrix $ A $ can be transformed into an upper-triangular matrix by simultaneously combining row and column transformations, we say that $ A $ is \emph{reducible}; otherwise, $ A $ is \emph{irreducible}.

\section{Preliminaries}\label{section2}

\subsection{Graph Theory}
A weighted, undirected graph is defined as $ \mathcal{G} (\mathcal{V} , \mathcal{E} , \boldsymbol{A}) $, which represents the whole nominal communication topology of $ N $ nodes.
For the three components in $ \mathcal{G} (\mathcal{V}, \mathcal{E}, \boldsymbol{A}) $, $ \mathcal{V} = \{v_1, v_2, \dots, v_n\} $ denotes a set of all nodes; $ \mathcal{E} = \left\lbrace E(i, j) :v_i, v_j \in \mathcal{V} \right\rbrace \subset \mathcal{V} \times\mathcal{V} $ denotes an undirected edge set, where $ E(i, j) $ represents one edge between nodes $ v_i $ and $ v_j $; and the adjacency matrix $ \boldsymbol{A} = [a_{i j}] \in \mathbb{R}^{N\times N}_{> 0} $ denotes the weight of $ \mathcal{G} $. 
$\mathcal{N}_{i}=\{v_{j}\in \mathcal{V}: E(i, j)\in \mathcal{E} \}$ is the neighbor set of node $ v_i $.
The node set of all other nodes in $\mathcal{V} $ except node $ i $ is denoted by $\mathcal{V}_{-i}$.
The edge set of the complete graph associated with $ \mathcal{G} $ is denoted by $\mathcal{C}$. Assume that graph $ \mathcal{G} $ has no self-loops, i.e., $ a_{ii} = 0, \forall i \in \textbf{I}[1, N] $.
The adjacency matrix $ \boldsymbol{A} $ is reducible if its associated graph $ \mathcal{G} $ is strongly connected.

\subsection{Game Theory}

Denote a group of $ N $ players as $ \mathcal{P} = \{P_1, P_2, \dots, P_N \} $. Each player makes its own strategy $ s_i \subset \mathbb{R}^m $. The conjoint decision set of all players is denoted by $ \mathcal{U} = \{ U_i \}_{i\in \textbf{I}[1,N]} \subset \mathbb{R}^{mN} $, then $ s_i\in U_i $. 
Denote that $ U_{-i} = U\, \backslash\, \{U_i \} \subset \mathbb{R}^{m(N-1)} $, which represents the set of all policy decisions except the $ i $th one. The policy profile of all players is denoted by $ s = (s_i, s_{-i}) \in \mathcal{U} $, where $s_{-i} \in U_{-i}$. Let $ J_i(s_i, s_{-i}): U_i \times U_{-i} \mapsto \mathbb{R}_{\geq 0} , i \in \textbf{I}[1,N] $, define the cost function of the $ i $th player, which is also relevant to the decisions of other players apart from itself. Then, a multiplayer game \cite{ye2017distributed} can be defined as
$ \mathds{G} (\mathcal{P}, \mathcal{U}, \mathcal{J}) $
with $\mathcal{J}=\{J_i(s_i, s_{-i})\}_{i\in \textbf{I}[1,N]}$.

\begin{myDef}\label{Y-def-1}
A policy profile $s^*=(s_i^*, s_{-i}^*)\in U $ can be considered as a Nash equilibrium of $ \mathds{G} (\mathcal{P}, \mathcal{U}, \mathcal{J}) $, if for all $s_i\in U_i$, it follows that
	\begin{equation*}
		J_i(s_i^*, s_{-i}^*)\leq J_i(s_i, s_{-i}^*),\; i\in \textbf{I}[1,N].
	\end{equation*}
At the Nash equilibrium, no player can obtain greater benefits by unilaterally altering its own strategy \cite{chen2018security,ye2017distributed}.
\end{myDef}

\subsection{Geometric Programming}
We first introduce the class of posynomial functions \cite{Stephen2007A}. Denote $ \eta_1, \dots , \eta_n $ as $ n $ positive variables and define $ \eta = (\eta_1, \dots , \eta_n)^\mathrm{T} \in \mathbb{R}^n_{> 0} $. If there exists a function satisfying  the form
$ h : \mathbb{R}^n_{> 0} \to \mathbb{R} = d\eta^{a_1}_1\eta^{a_2}_2 \dots\ \eta^{a_n}_n $ with $ d \in \mathbb{R}_{> 0} $ and $ a_i\in \mathbb{R} $, $ \forall i \in \textbf{I}[1,n] $, we say it is a \emph{monomial}. Moreover, the sum of monomials is called a \emph{posynomial}, that is, $ f : \mathbb{R}^m_{> 0} \to \mathbb{R} =\sum^M_{m = 1} c_mh_m(\eta)  $ with $ c_m \in  \mathbb{R}_{> 0} $ and $ h_m(\eta) = d_m\eta^{a_{1, m}}_1\eta^{a_{2, m}}_2 \dots  \eta^{a_{n, m}}_n $ for all $ m $. Given monomial functions $h_1, \dots , h_q$ and posynomial functions $f_0, \dots, f_p$, a GP problem can be described as
	\begin{align}
		\min \limits_{\eta \in \mathbb{R}^n_{> 0}} \quad &f_0 (\eta) \nonumber \\
		{\rm s.t.~} \quad &f_i(\eta) \leq 1, ~~\forall\; i \in \textbf{I}[1,p],  \\
		&h_j(\eta) = 1, ~~\forall\; j \in \textbf{I}[1, q],\nonumber
	\end{align}
where $ f_i(\eta)\leq 1, \forall i \in \textbf{I}[1, p] $, denote posynomial constraints. The GP problem can be converted into a convex optimization problem by a logarithmic variable transformation, that is, $ \eta = \exp [ \textbf{x} ], \textbf{x} \in \mathbb{R}^n $. 
	
\begin{myDef}[{\cite{preciado2014optimal}}]\label{Y-def-2}
For a posynomial $ f : \mathbb{R}^n \to \mathbb{R} $, the log-scale function is defined as
	\begin{equation} \nonumber
		F(\textbf{x}) = \log( f (\exp[\textbf{x}])),
	\end{equation}
which is convex in $ \textbf{x} $.
\end{myDef}

Thus, GP problem (1) can be rewritten in the following standard form \cite{boyd2004convex}:
	\begin{align}\label{SG}
		\min \limits_{\textbf{x} \in \mathbb{R}^n} \quad &F_0 (\textbf{x}) \nonumber \\
		{\rm s.t.~} \quad &F_i(\textbf{x}) \leq 1, ~~\forall\; i \in \textbf{I}[1,p],  \\
		&H_j(\textbf{x})=0, ~~\forall j\in \textbf{I}[1,q],\nonumber
	\end{align}
where $F_0(\textbf{x})$ and $F_i(\textbf{x})$ are defined in Definition \ref{Y-def-2}.
For a monomial $h_j(\textbf{x})=d_jx_1^{a_{1, j}}x_2^{a_{2, j}}\dots x_n^{a_{n, j}}$, one can obtain that $ H_j(\textbf{x})= b_j^{\mathrm{T}}(\textbf{x})+\log d_j $ with $b_j= (a_{1, j},a_{2, j},\dots, a_{n, j})^{\mathrm{T}}$.

	
\subsection{Some Useful Results and Lemmas on Positive Systems}
Consider the following two interconnected discrete-time positive systems \cite{farina2000positive}: 
	\begin{equation}\label{s1}
		\mathds{S}_{1}:
		\begin{cases}
			x_{1}( k+1 )=A_{1}x_{1}(k)+B_{1}u_{1}(k), \\
			y_{1}( k )=C_{1}x_{1}(k)+D_{1}u_{1}(k),
		\end{cases}
	\end{equation}
	and
	\begin{equation}\label{s2}
		\mathds{S}_{2}:
		\begin{cases}
			x_{2}( k+1 )=A_{2}x_{2}(k)+B_{2}u_{2}(k), \\
			y_{2}( k )=C_{2}x_{2}(k)+D_{2}u_{2}(k),
		\end{cases}
	\end{equation}
where $x_i \in \mathbb{R}^n$, $u_i \in \mathbb{R}^m$, $y_i \in \mathbb{R}^q$, $i \in \{1,2\}$, are respectively the system state and $ A_i \in \mathbb{R}^{n\times n}$, $ B_i \in \mathbb{R}^{n\times m}$, $ C_i \in \mathbb{R}^{q\times n}$, $ D_i \in \mathbb{R}^{q\times m}$, $i \in \{1,2\}$, are nonnegative matrices. Assume that the above systems both have compatible dimensions, that is, $ y_2= u_1 $ and $ y_1 = u_2 $. The interconnected system that consists of (\ref{s1}) and (\ref{s2}) is denoted by $ \mathds{S}_{1\oplus 2} $. Furthermore, an assumption is introduced to guarantee that $ \mathds{S}_{1\oplus 2} $ is well-defined:
	
	\begin{assumption}\label{Y-ass-1}
$ \mathds{S}_{1\oplus 2} $ is well-posed, which means $ D_1D_2 $ is Schur stable.
	\end{assumption}

Under Assumption \ref{Y-ass-1}, the necessary and sufficient conditions to prove the stability of $ \mathds{S}_{1\oplus 2} $ are as follows.

    \begin{lemma}\label{Y-lem-1}
$ \mathds{S}_{1\oplus 2} $ is Schur stable, if and only if there exist vectors $  h_i $, $ q_i > 0 $, $i \in \{1,2\}$ such that
\begin{equation}
	\begin{cases}
		h_1^\mathrm{T} (A_1 - I)+q_1^\mathrm{T} C_1 < 0, \\
		h_1^\mathrm{T} B_1+q_1^\mathrm{T} D_1-q_2^\mathrm{T} <0, \\
		h_2^\mathrm{T} (A_2 - I)+q_2^\mathrm{T} C_2 < 0, \\
		h_2^\mathrm{T} B_2+q_2^\mathrm{T} D_2-q_1^\mathrm{T} <0.
	\end{cases}
\end{equation}
\end{lemma}	

\textbf{Proof.}
Based on (\ref{s1}), (\ref{s2}) and Assumption \ref{Y-ass-1}, it follows that
\begin{small}
\begin{align*}
  \begin{bmatrix}
   x_1(k+1)\\
   x_2(k+1)
  \end{bmatrix}
  & \!\!= \!\!\begin{bmatrix}
   A_1&0\\
   0&A_2
\end{bmatrix}\!\!
\begin{bmatrix}
   x_1(k)\\
   x_2(k)
\end{bmatrix}\\
&\!\!- \!\! \begin{bmatrix}
   0&B_1\\
   B_2&0
\end{bmatrix}
\!\! \begin{bmatrix}
   -I&D_1\\
   D_2&-I
\end{bmatrix}^{-1}
\!\! \begin{bmatrix}
   C_1&0\\
   0&C_2
\end{bmatrix}
\!\! \begin{bmatrix}
   x_1(k)\\
   x_2(k)
\end{bmatrix} \!,
\end{align*}
\end{small}
is internally stable. Thus, the matrix
\begin{small}
\begin{equation*}
\begin{bmatrix}
   A_1-I&0\\
   0&A_2-I
\end{bmatrix} \!\! - \!\! \begin{bmatrix}
   0&B_1\\
   B_2&0
\end{bmatrix}\!\!\begin{bmatrix}
   -I&D_1\\
   D_2&-I
\end{bmatrix}^{-1}\!\!\begin{bmatrix}
   C_1&0\\
   0&C_2
\end{bmatrix}\!,
\end{equation*}
\end{small}
is a Hurwitz matrix. When the definition of the Schur complement \cite{zhang2006schur} is recalled, it can be proven that the matrix
\begin{small}
\begin{equation*}
\mathds{A}=
\begin{bmatrix}
  A_1-I&0&0&B_1\\
   0&A_2-I&B_2&0\\
   C_1&0&-I&D_1\\
   0&C_2&D_2&-I
\end{bmatrix}\!,
\end{equation*}
\end{small}
is a Hurwitz matrix. Based on Assumption \ref{Y-ass-1} and \cite[Lemma 4.4]{rami2012characterization}, the Hurwitz stability of matrix $\mathds{A}$ is equivalent to stating that there exist vectors $h_i, q_i>0$, $i=1,2$, such that
\begin{small}
\begin{equation*}
\begin{bmatrix}
   h_1\\
   h_2\\
   q_1\\
   q_2
\end{bmatrix}^{\mathrm{T}}
\begin{bmatrix}
   A_1-I&0&0&B_1\\
   0&A_2-I&B_2&0\\
   C_1&0&-I&D_1\\
   0&C_2&D_2&-I
\end{bmatrix}<0.
\end{equation*}
\end{small}
Since $\mathds{A}$ is a Metzler matrix, Lemma \ref{Y-lem-1} is proven. 
$\hfill \hfill \blacksquare $
\vspace{0.2cm}

	\begin{lemma}[\rm{\cite[Lemma 4.3]{ogura2019geometric}}]\label{Y-lem-2}
Let $ F $ denote a nonnegative matrix and $ \gamma $ denote a positive scalar. The following two conditions are equivalent:
	\begin{enumerate}
	\item $ \| F \| < \gamma $.
	\item There exist positive vectors $ u, v $, such that
    \end{enumerate}
	\begin{equation} \nonumber
	Fu<\gamma v, ~ F^\mathrm{T} v < \gamma u.
	\end{equation}
	\end{lemma}

	\begin{lemma}[\rm{\cite[Lemma 1]{briat2013robust1}}]\label{Y-lem-3}
Let $ g\in \mathbb{R}^n $ and $ H\in \mathbb{R}^{m\times n} $ be a nonnegative vector and matrix, respectively. Assume that $ M \in \mathbb{R}^{n\times n} $ is a Metzler matrix and $ v \in \mathbb{R}^m $. The following conditions are equivalent:
	\begin{enumerate}
	\item $ M $ is Hurwitz with $ -HM^{-1}g < v $.
	\item There exists a vector $ \omega\in \mathbb{R}^n_{> 0} $, such that $ H\omega < v $ and $ M\omega + g < 0 $.
    \end{enumerate}
	\end{lemma}

\subsection{Overview of the Foschini-Miljanic Algorithm}
We briefly review the FM Algorithm by borrowing the notation in \cite{2016A}. Suppose that there exist $ N $ communication channels, each channel corresponding to a receiver-transmitter pair. The transmission gain for the $ i $th channel and the interference gain from the $ i $th channel to the $ j $th channel are denoted by $ h_i \in (0, 1] $ and
$ g_{ji} $, respectively. The variance of the thermal noise at the $ i $th receiver is denoted by $\nu_i$. The power level chosen by the $ i $th transmitter is denoted by $ p_i $. 
The transmission quality of the $ i $th channel is measured by the signal-to-interference-plus-noise-ratio (SINR), which is defined as
	\begin{equation} \nonumber
		{\rm SINR}:\gamma_{i}=\frac{h_ip_i}{\nu_i+\sum_{j \neq i}g_{ji}p_j}, \; i, j\in \mathbf{I}{[1,N]}.
	\end{equation}
Suppose that the $ i $th receiver can measure its $ \gamma_{i} $ and transfer it to the transmitter.
Each user regulates its power level through the following FM Algorithm \cite{1993A}:
\begin{flalign}\label{FM1}
	& p_i(k+1) \! - \! p_i(k) \!=\! k_i \Big(\!\! -p_i(k) \!+\! \frac{\gamma_i}{h_i} \big( \nu_i
	\!+ \!\! \sum\nolimits_{j \neq i} g_{ji}p_j(k) \big) \Big),& 
\end{flalign}
where $ k_i \in \mathbb{R}_{> 0} $ is a proportionality constant.
It was proven in \cite{1993A} that if there exists a vector $ \bar{p} = (\bar{p}_1, \dots, \bar{p}_N)^\mathrm{T} $ such that 
\begin{equation} \label{FM32-2}
	\frac{h_i\bar{p}_i}{\nu_i+\sum_{j \neq i}g_{ji}\bar{p}_j} \geq \bar\gamma_i, 
	~~\forall i\in \mathbf{I}{[1,N]},
\end{equation} 
where $ \bar\gamma_i \in \mathbb{R}_{\geq 0} $ is a constant and denotes the predefined SINR of the $ i $th channel which guarantees the QoS of users, then Eq. (\ref{FM1}) will converge to the solution $\bar{p}_i$. It is also the purpose of the FM algorithm, that is, to find the set of power levels $(\bar{p}_1, \dots, \bar{p}_N)$ to satisfy the inequality (\ref{FM32-2}).

	
Define $ p = \left[ p_1, \ldots, p_N\right]^\mathrm{T} $ and $ \nu =\left[ \nu_1, \ldots, \nu_N\right]^\mathrm{T} $. Additionally, define the matrices $ K = {\rm diag}\left( k_1, \ldots  , k_N\right)  $, $ \bar{\varGamma} = {\rm diag}(\bar\gamma_1, \ldots  , \bar\gamma_N ) $, $  H = {\rm diag}(h_1, \ldots  , h_N)$, and the interference matrix $ G $ such that
	\begin{equation} \nonumber
		G_{ji}=
		\begin{cases}
			0,& \text{if} ~~ i\, =\, j,\\
			g_{ji},& \text{otherwise.}
		\end{cases}
	\end{equation}
Eq. (\ref{FM1}) can be rewritten in the following compact form:
	\begin{equation} \label{FM32}
		P(k+1)\!=\!(K\left(-I_N+\bar{\varGamma}H^{-1}G\right)+I_N)P(k)+K \bar{\varGamma} H^{-1} \nu,
	\end{equation}
since $ K\left( -I_N+\bar{\varGamma}H^{-1}G\right) $ is Metzler and $ K \bar{\varGamma} H^{-1} $ is nonnegative, (\ref{FM32}) is obviously a positive system.
As shown in \cite{2012Unconditional, 2016A}, the stability analysis of the FM algorithm reduces to the stability analysis of the following system:
	\begin{equation}\label{FM2}
		X(k+1)=(K\left(-I_N+\bar{\varGamma}H^{-1}G\right)+I_N) X(k),
	\end{equation}
with the origin being its unique equilibrium.

\section{Main Results}\label{section3}

In this section, we first develop a GoG framework. 
Second, it is proven that the robust stabilization problem of the discrete-time FM algorithm against the bounded structured uncertainties of the 1-norm and 2-norm can be transformed into a standard GP problem, as shown in Lemma \ref{Y-the-1}. Since the existence of the attacker can be regarded as the worst case of structured uncertainties, the above GP framework can be utilized as a basis to solve the problem of interference suppression under the threat of malicious attacks.
Third, it is verified that the GP scheme can be extended to determine the GNE for our GoG.
Fourth, a heuristic algorithm is established in a round-robin manner, where the optimal strategy of two subnetwork policymakers is computed by iterative applications of GP.
Furthermore, the attack mode and capability, as well as the greedy heuristic strategy under malicious adding-edge attacks, are investigated.

\subsection{A GoG Framework of the Foschini–Miljanic Algorithm}

\begin{figure}[!]
  \centering
  \includegraphics[width=0.485\textwidth]{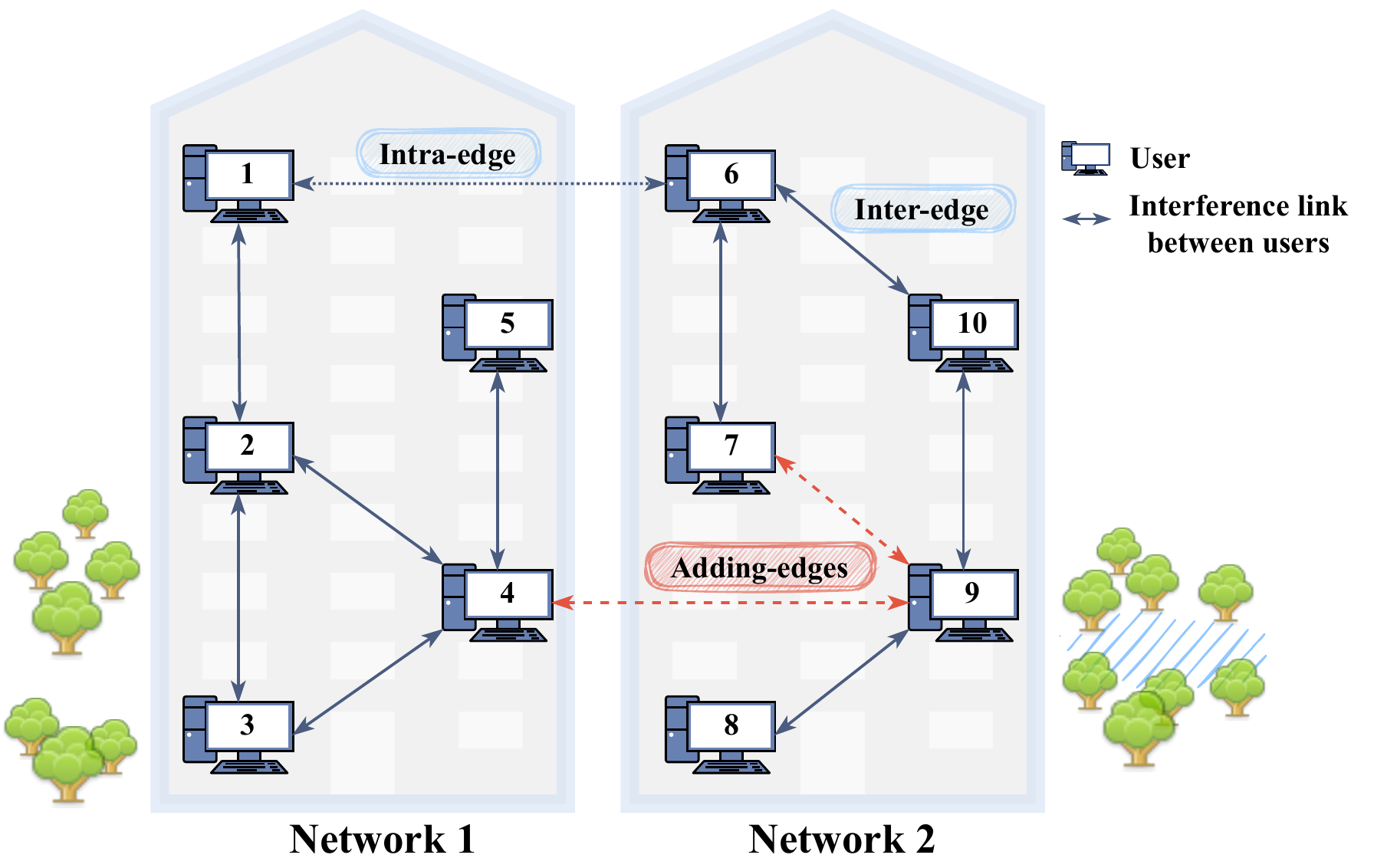}
  \caption{An example of a complete network with different types of edges.}
  \label{fig:figure1}
\end{figure}

Denote the two subnetworks as Network $1$:
$ \mathcal{G}_1=(\mathcal{V}_1,\mathcal{E}_1, \boldsymbol{A}_{11}) $, where the set of nodes $ \mathcal{V}_1=\{1,2,\dotsc,M\} $, the set of inter-edges $ \mathcal{E}_1 $ are among nodes in the same subnetwork,
and the adjacency matrix $ \boldsymbol{A}_{11} $;
and Network 2: $ \mathcal{G}_2=(\mathcal{V}_2,\mathcal{E}_2, \boldsymbol{A}_{22}) $, where the set of nodes $ \mathcal{V}_2=\{M+1,M+2,\dotsc,N\}$, the set of inter-edges $ \mathcal{E}_2 $, and the adjacency matrix $ \boldsymbol{A}_{22} $. In the whole topology $\mathcal{G}=(\mathcal{V},\mathcal{E}, \boldsymbol{A})$, it follows that
\begin{equation*}
  \boldsymbol{A}=
  \begin{bmatrix}
    \boldsymbol{A}_{11} &   \boldsymbol{A}_{12} \\
    \boldsymbol{A}_{12}^{\mathrm{T}} &   \boldsymbol{A}_{22}
    \end{bmatrix},
\end{equation*}
$ \boldsymbol{A}_{12} $ denotes the adjacency matrix corresponding to intra-edges, that is, the edges between the two subnetworks. Let the nodes connected directly to another subnetwork denote as border nodes, such as nodes $1$ and $6$ in Fig. \ref{fig:figure1}.

The common goal of the two subnetwork policymakers is to improve the overall performance of communication networks. They adopt a non-fully cooperative manner to allocate communication resources; specifically, both make decisions without anticipating the current policies of the other.
We further investigate the case where a malicious attacker attempts to enhance interference among the channels by adding several key edges to the topology ${\mathcal{G}}$.
The compromised network topology and its adjacency matrix are defined as $\tilde{\mathcal{G}}$ and $\tilde{\boldsymbol{A}}=\boldsymbol{A}+\boldsymbol{A}_q$, respectively, where $\boldsymbol{A}_q \subset \mathcal{C}\setminus \mathcal{E}$ is the alternating adjacency matrix induced by the attacker.
Assume that the attack magnitudes $ q_1=\|\boldsymbol{A}_q\|_{1} $ and $ q_2=\|\boldsymbol{A}_q\|_{2} $ satisfy the constraints $ q_1\leq \bar{q}_1 $ and $ q_2\leq \bar{q}_2 $, respectively, where $ \bar{q}_1 $ and $ \bar{q}_2 $ are two calculated known positive constants.

\begin{remark}
The description of attacking magnitudes aims to avoid a trivial conclusion: 1) Considering that the attacker is hidden in the dark and cannot be detected, its attacks are different from the usual uncertainties or interference. Thus, the constraint $ q_1\leq \bar{q}_1 $ indicates that the attacker can add up to $ \bar{q}_1 $ edges originating from one node.
2) Considering that the ability of the attacker is restricted by factors such as geographical location and device performance, constraint $ q_2\leq \bar{q}_2 $ indicates that there exists an upper bound on the expansion of network connectivity.
Notably, the unknown $ \bar{q}_2 $ will prevent us from seeking the exact GNE. Hence, the given values of $ \bar{q}_1 $ and $ \bar{q}_2 $ are indispensable in this work.
In addition, the allowable upper bound of $ \bar{q}_2 $ can be calculated via Theorem \ref{Y-the-3}.
$\hfill \hfill \square $
\end{remark}

\begin{figure}[!]
  \centering
  \includegraphics[width=0.48\textwidth]{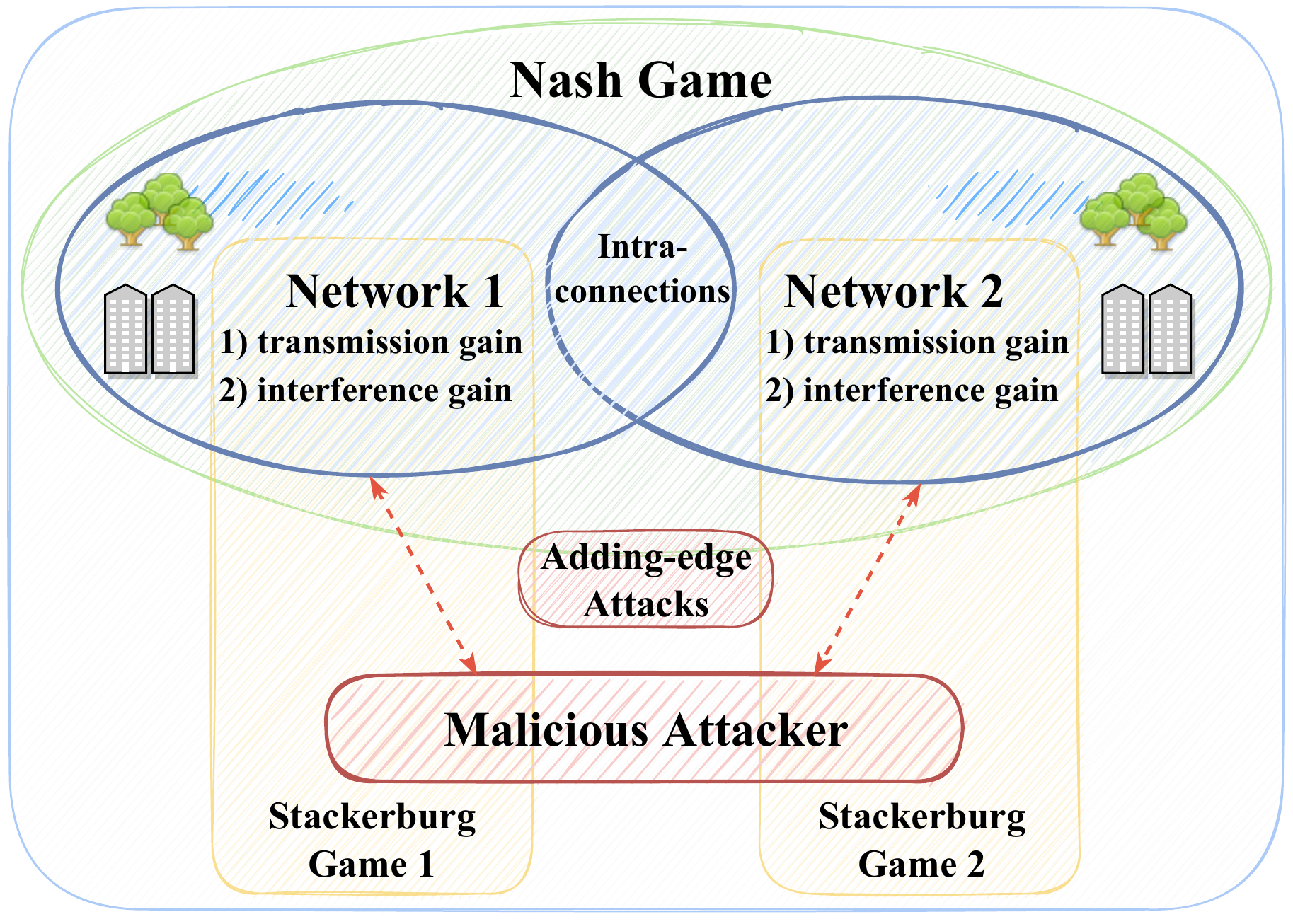}
  \caption{A GoG (Gestalt game) framework.}
  \label{fig:figure2}
\end{figure}
In the scenario of the two subnetwork policymakers versus an adding-edge attacker, the process of guaranteeing the QoS of the entire network, that is, each node reaching the required SINR can be described as the GoG framework shown in Fig. \ref{fig:figure2}:
\begin{enumerate}
  \item The first Stackelberg game (SG) \cite{maharjan2013dependable} indexed by 1 consists of the policymaker in Network 1 and the attacker, who make decisions in turn;
  \item Similarly, the second Stackelberg game indexed by 2 involves the policymaker in Network 2 and the attacker;
  \item The two policymakers constitute the player set of a Nash game (NG) and make decisions without anticipating the current policy of the other.
\end{enumerate}

The introduction of antenna array technology, such as beamforming \cite{6736761}, can effectively increase the transmission gain and reduce the interference gain.
The measures taken will incur the corresponding costs, while the budget is limited.
Therefore, how to allocate certain communication resources to realize the stability of the FM algorithm is a problem worth discussing.
Assume that interference will occur in the $ i $th and $ j $th channels if the $ i $th and $ j $th nodes are adjacent in network $ \mathcal{G} $, and the interference gain is denoted by $ g_{ji} $.
For tuning the interference gain and transmission gain to be $ g_{ji} $ and $ h_i $, respectively, the relevant unit costs are $ \alpha(g_{ji}) $ and
$ \beta(h_i) $. Moreover, due to resource limitations, we suppose the gains can be tuned in the following intervals:
	\begin{equation}\label{Cons1}
		\begin{gathered}
			0< \underline{h} \leq h_i \leq \bar{h}, ~~\forall i\in \mathbf{I}{[1,N]},\\
			0< \underline{g} \leq g_{ji} \leq \bar{g}, ~~\forall i,j\in \mathbf{I}{[1,N]} \, \rm{and} \, j \neq i.
		\end{gathered}
	\end{equation}

Then, the total cost for realizing the performance is
	\begin{equation} \nonumber
		L(g, h) =\sum_{i=1}^N \alpha(g_{ji}) +\sum_{i=1}^N \beta(h_i),
	\end{equation}
	where $ g, h \in \mathbb{R}^N $ are two column vectors obtained by superimposing $ g_{ji} $ and $ h_i $.

Let us introduce the vectorial parameters
\begin{align*}
&\theta_1=[~g_1,\dotsc,g_M,h_1,\dotsc,h_M]^{\mathrm{T}},\\
&\theta_2=[~g_{M+1},\dotsc,g_N,h_{M+1},\dotsc,h_N]^{\mathrm{T}},
\end{align*}
which represent the performances of Network 1 and Network 2, respectively.
Then define the total cost function of Network 1 for realizing $ \theta_1 $ as
\begin{equation*}\label{totalcost1}
  L_1(\theta_1)=\sum_{i=1}^{M}L_1^{i}(g_i,h_i)=
  \sum_{i=1}^{M}(\alpha(g_i)+\beta(h_i)),
\end{equation*}
Likewise, the total cost function of Network 2 is
\begin{equation*}\label{totalcost2}
  L_2(\theta_2)= \!\!\!\! \sum_{j=M+1}^{N} \!\!\! L_2^{j}(g_j,h_j) =
  \!\!\!\! \sum_{j=M+1}^{N} \!\!\! (\alpha(g_j)+\beta(h_j)).
\end{equation*}
Both policymakers aim to achieve optimal overall communication quality with a limited budget by tuning the gains.



Based on the above discussions, we propose the main problem investigated in this work:

\noindent \textbf{Problem GSA} (GNE Seeking under malicious Attacks):
Finding an accurate GNE of the Gestalt game is equivalent to determining the policy profiles of the two subnetwork policymakers and the strategy of the external malicious adding-edge attacker, that is,
$(\{\theta_i^*\}_{i\in [1,2]}, \boldsymbol{A}_q^*)$,
such that
\begin{enumerate}
  \item $(\theta_{\pi}^*, \boldsymbol{A}_q^*)$, $ \pi\in [1,2]$, is a Stackelberg equilibrium;
  \item ($\theta_1^*$, $ \theta_2^* $) is a Nash equilibrium;
  \item the attacking magnitudes satisfy $ \|\boldsymbol{A}_q^*\|_1 \leq \bar{q}_1,~\|\boldsymbol{A}_q^*\|_2\leq \bar{q}_2 $;
  \item the total cost function
  $L_{\rm sum}(\theta)=L_1(\theta_1)+L_2(\theta_2)$ can be optimized.
\end{enumerate}

Next, we provide several valuable theoretical results for finding the solution to Problem \textbf{GSA}.

\subsection{Robust Stabilization against Norm-bounded Structured Uncertainties}
Referring to the stability analysis of robust stabilization problem under structured uncertainties in \cite{colombino2015convex}, the interference matrix $ G $, which is not completely known, can be described in the following parameter form
	\begin{equation} \nonumber
		G + E \Delta F,
	\end{equation}
where $ G $ is a nominal interference matrix, $ \Delta \subset \bm\Delta \in \mathbb{R}^{m\times m} $ is an unknown nonnegative matrix, and $ E \in \mathbb{R}^{N\times m} $ together with $ F \in \mathbb{R}^{m\times N} $ are nonnegative matrices. Without loss of generality, we introduce the following assumption for simplicity:
\begin{assumption}\label{Y-ass-2}
The matrices $ E $, $ F $ and $ \Delta $ have the same dimensions in this paper, that is, $ m = N $.
\end{assumption}
	
We now focus on solving the robust stabilization problem of the FM algorithm against nonnegative structured uncertainties via GP with finite decision variables. Specifically, the uncertainty set $ \bm\Delta $  \cite{colombino2015convex} is defined as
	\begin{align*} \nonumber
	\bm\Delta = \big\{
    &\text{blkdiag}
(\Delta_1,\dots,\Delta_\phi,\delta_{\phi+1},\dots,\delta_{\phi+\sigma}): \\
	& \Delta_k \in \mathbb{R}^{m_k \times m_k}_{\geq 0}, k=1, \dots, \phi, \\
	& \delta_k \in \mathbb{R}_{\geq 0},k=\phi+1,\dots, \phi+\sigma \big\}
    \subset \mathbb{R}^{N\times N}_{\geq 0}.
	\end{align*}
	
\begin{figure}[!]
  \centering
  \subfigure[Two interconnected positive systems]{
  \begin{minipage}[b]{0.227\textwidth}
  \includegraphics[width=1\textwidth]{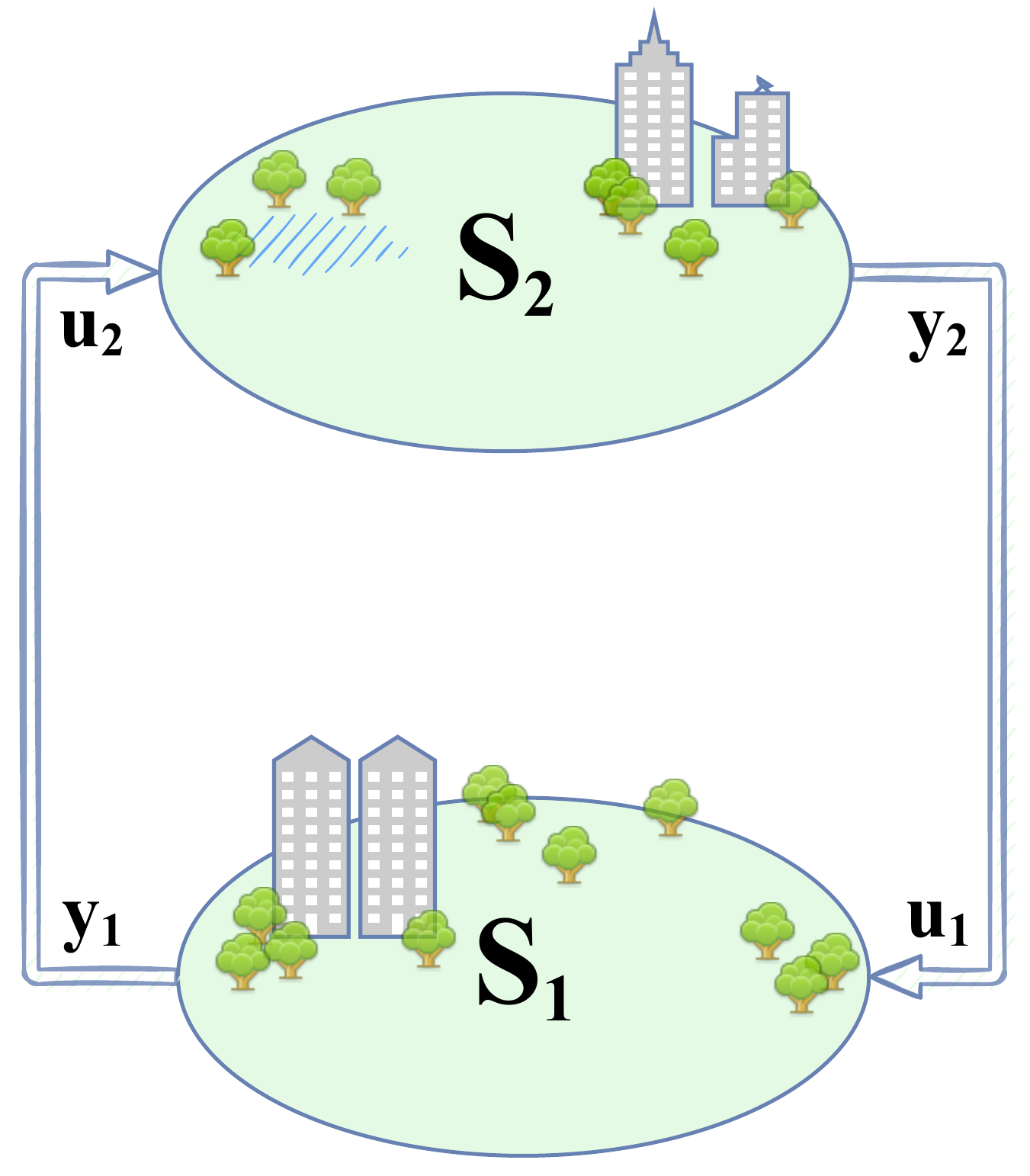}
  \end{minipage}
  }
  \subfigure[The positive system with uncertainties ]{
  \begin{minipage}[b]{0.227\textwidth}
  \includegraphics[width=1\textwidth]{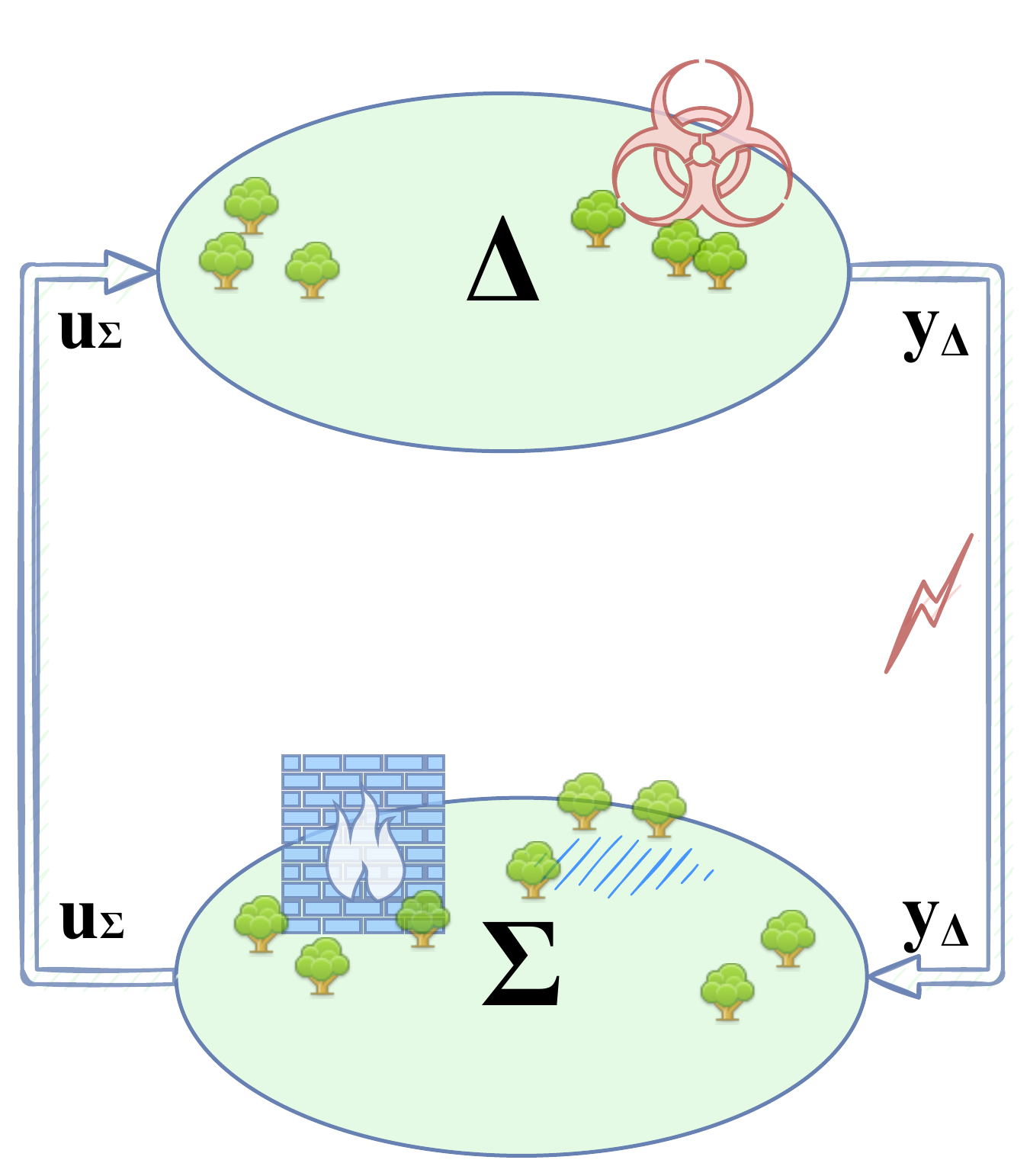}
  \end{minipage}
  }
  \caption{A demonstration of applying $\mathds{S}_{1\oplus2}$ to $\mathds{S}_{\Sigma\oplus\Delta}$.}
  \label{fig:figure0}
  \end{figure}

Considering the discrete-time positive linear system based on the FM algorithm:
\begin{align}\label{X10}
	x(k+1) =& (K(-I_N+\bar{\varGamma}H^{-1}G) + I_N)x(k) \nonumber\\
	&+K\bar{\varGamma} H^{-1}E\Delta F x(k).
\end{align}
The above closed-loop system can also be regarded as the open-loop system $\Sigma$ 
\begin{flalign} \label{S10}
&\Sigma:
\left\{
\begin{aligned}
&x(k+1)=
\begin{aligned}[t]
&(K(-I_N+\bar{\varGamma}H^{-1}G) + I_N)x(k) \\
& ~+ K\bar{\varGamma} H^{-1}E w(k), \\
\end{aligned}\\
&y(k)= 
\begin{aligned}[t]
&F x(k),\\
\end{aligned}\\
\end{aligned}
\right.&
\end{flalign}
where $ \Delta\in \bm\Delta\subset \mathbb{R}^{N \times N}_{\geq 0} $denotes an uncertain but constant matrix, closed by the following relationship
\begin{equation}\label{U1}
	u = \Delta y.
\end{equation}
Then, we concentrate on the robust stability of $ \Sigma $ aroused from the uncertain matrix $ \Delta $, which is related to the internal stability of the system (\ref{X10}).
The above closed-loop system is denoted as $ \mathds{S}_{\Sigma\oplus \Delta} $. As shown in Fig. \ref{fig:figure0}, when choosing
$ A_1 = K(-I_N+\bar{\varGamma}H^{-1}G) + I_N $,
$ B_1 = K\bar{\varGamma} H^{-1}E $,
$ C_1 = F $, $ B_2 = \Delta $, $ C_2 = I_N$, and
$ A_2 = D_1 = D_2 = \boldsymbol{0}_N $,
system $ \mathds{S}_{\Sigma\oplus \Delta} $ is a case of system $ \mathds{S}_{1\oplus2} $.

\begin{lemma}\label{Y-lem-4}
$\mathds{S}_{\Sigma\oplus\Delta} $ is robust stable, if there exist vectors $ \rho, q_1, q_2\in \mathbb{R}^N_{> 0} $ such that
	\begin{equation}
		\begin{cases}
		\rho^\mathrm{T}K(-I_N+\bar{\varGamma}H^{-1}G)+q_1^\mathrm{T}F<0,\\
		\rho^\mathrm{T}K\bar{\varGamma} H^{-1}E-q_2^\mathrm{T}<0,\\
		q_2^\mathrm{T}\Delta-q_1^\mathrm{T} \leq 0.
		\end{cases}
	\end{equation}	
\end{lemma}	

\textbf{Proof.}
Assumption \ref{Y-ass-1} is satisfied in our topic because of $D_1 = D_2 = \boldsymbol{0}_N$. The conditions to guarantee the robust stability of $\mathds{S}_{\Sigma\oplus\Delta} $ in Lemma \ref{Y-lem-4} can be easily obtained by recalling Lemma \ref{Y-lem-1}.
$\hfill \hfill \blacksquare $
\vspace{0.2cm}



	
Lemma \ref{Y-lem-4} is utilized to prove the robust stability of $ \mathds{S}_{\Sigma\oplus\Delta} $ under unit 1-norm-bounded structured uncertainty.
Next, we introduce the following conditions:
\begin{equation}\label{Y-FM1}
\left\{
\begin{aligned}
\varGamma_1 = & ~ {\rm sup}_{\Delta\in\bm\Delta, \| \Delta \|_1\leq \varepsilon_1}\lambda_{max}(K(-I_N+\bar{\varGamma}H^{-1}G)\\
&+K \bar{\varGamma} H^{-1}E \Delta F)<0,\\
\varGamma_2 = & ~ {\rm sup}_{\Delta\in\bm\Delta, \| \Delta \|_2\leq \varepsilon_2}\lambda_{max}(K(-I_N+\bar{\varGamma}H^{-1}G)\\
&+K \bar{\varGamma} H^{-1}E \Delta F)<0,\\
\end{aligned}
\right.
\end{equation}
where $ \varepsilon_i\in \mathbb{R}_{> 0} $, $ i \in \{1, 2\} $, represent the upper bounds of the 1-norm and 2-norm of $ \Delta $, respectively. To transform the inequality constraints from being strict to nonstrict, two small positive scalars $ \varsigma_1, \varsigma_2 \in (0, 1) $ are introduced:
\begin{align}
		& \varGamma_1\leq-\varsigma_1, \label{T1} \\
		& \varGamma_2\leq-\varsigma_2. \label{T2}
\end{align}

Define the vectorial decision variable $\theta=[\theta_1^{\mathrm{T}},\theta_2^{\mathrm{T}}]^{\mathrm{T}}$. The total cost function $L(\theta)$ satisfies Assumption \ref{Y-ass-4} with  $L_0=N(\frac{\bar{g}^{-p}}{\underline{g}^{-p}-\bar{g}^{-p}}+
\frac{\underline{h}^{q}}{\bar{h}^{q}-\underline{h}^{q}})$.
Next, we investigate the following robust stabilization problem:
\begin{align}\label{Problem10}
\mathds{P}_{1}:~~
\min_{\theta \in \Theta} ~~~ &L(\theta) \\
\rm s.t.~~~~ &(\ref{T1})~{\rm and}~ (\ref{T2}), \nonumber
\end{align}
where $\varsigma_1, \varsigma_2$ in (\ref{T1}) and (\ref{T2}) represent the asymptotically stable rate of $ \mathds{S}_{\Sigma\oplus \Delta} $.
Next, we give the equivalent conditions of (\ref{T1}) and (\ref{T2}), respectively.
\begin{lemma}\label{Y-pro-2}
For system $ \mathds{S}_{\Sigma\oplus \Delta} $, stating that the inequality (\ref{T1}) holds is equivalent to stating that there exists a positive vector $ \rho\in \mathbb{R}^N_{> 0} $, such that the following inequalities hold:
\begin{equation}\label{C1}
\left\{
\begin{split}
&(K(-I_N+\bar{\varGamma}H^{-1}G) + \varsigma_1 I_N)^\mathrm{T} \rho + \sqrt{\varepsilon_1}F^\mathrm{T} \boldsymbol{1}_N <0, \\
&\sqrt{\varepsilon_1}E^\mathrm{T} H^{-1} \bar{\varGamma} K \rho-\boldsymbol{1}_N<0.
\end{split}
\right.
\end{equation}
\end{lemma}	

\textbf{Proof.} If Eq. (\ref{T1}) holds, the following  system
\begin{flalign}
&\mathds{S}^1_\varsigma:
\left\{
\begin{aligned} \nonumber
&x(k+1)=
\begin{aligned}[t]
&(K(-I_N+\bar{\varGamma}H^{-1}G)+ I_N +\varsigma_1 I_N)x(k)\\
&+ \sqrt{\varepsilon_1}K\bar{\varGamma} H^{-1}E u(k),\\
\end{aligned}\\
&y(k) = 
\begin{aligned}[t]
&\sqrt{\varepsilon_1} F x(k),\\
\end{aligned}\\
\end{aligned}
\right.&
\end{flalign}
with feedback (\ref{U1}) is internally stable, and $ \forall \Delta\in \bm\Delta $ satisfying $ \| \Delta \|_1\leq 1 $.
The proof of sufficiency is obvious by selecting $ q_1 = q_2 = \boldsymbol{1}_N $ in Lemma \ref{Y-lem-4}.
Then, \emph{Reduction to Absurdity} is employed to prove the necessity part.
Assume that condition (\ref{C1}) does not hold and denote the transfer function of $ \mathds{S}^1_\varsigma $ as 
\begin{equation}\nonumber
    M_{\Sigma}(z) \! = \! F (zI - K(-I_N+\bar{\varGamma}H^{-1}G) - \varsigma_1 I_N )^{-1} K\bar{\varGamma} H^{-1}E,
\end{equation}
there at least one index $j^\star$ according to \cite[Theorem 1]{ebihara2016analysis}, such that
\begin{equation*}
    [\boldsymbol{1}_N^{\mathrm{T}}M_{\Sigma}(0)]_{j^\star}\geq [\boldsymbol{1}_N]_{j^\star}=1.
\end{equation*}
Define $\Delta^{\star}=\textbf{e}_{j^\star}\boldsymbol{1}_N^{\mathrm{T}}$, where $\textbf{e}_{j^\star}$ is a vector of almost zero except the ${j^\star}$th element is equal to $1$. Furthermore, $\|\Delta^{\star}\|_{1}\in \boldsymbol{\Delta}\subset \mathbb{R}_{\geq 0}^{m\times r}$ and $\|\Delta^{\star}\|_{1}\leq 1$ satisfy the requirement on $\Delta$ in Lemma \ref{Y-pro-2}. It follows that
\begin{equation*}
\boldsymbol{1}_N^{\mathrm{T}}M_{\Sigma}(0)\Delta^{\star}=[\boldsymbol{1}_N^{\mathrm{T}}
M_{\Sigma}(0)]_{j^\star}\boldsymbol{1}_N^{\mathrm{T}}\geq \boldsymbol{1}_N^{\mathrm{T}},
\end{equation*}
which implies that $\mathds{S}_{\Sigma\oplus \Delta^\star}$ is not Schur stable because of its transfer function  $M_{\Sigma \oplus \Delta^\star}(0) = M_{\Sigma}(0) M_{\Delta^\star}(0) = M_{\Sigma}(0) \Delta^{\star}$. This is obviously contradictory since the robust stability condition is violated.
Therefore, one can obtain that $ \mathds{S}^1_\varsigma $ is robust stable, for all $ \Delta\in \bm\Delta $ satisfying $ \| \Delta \|_1\leq 1 $, if and only if there exists $\rho\in \mathbb{R}_{> 0}^N$ such that the inequalities in (\ref{C1}) hold.
$\hfill \hfill \blacksquare $
\vspace{0.2cm}

The following lemma shows that the robust stabilization problem can be reduced into a GP problem with finite variables.
\begin{lemma}\label{Y-pro-3}
Define the set
\begin{equation} \nonumber
\begin{split}
	\mathds{K} = \big\{ & \text{diag}(\pi_1 I_{m_1},\dots,\pi_\phi I_{m_\phi},\pi_{\phi+1},\dots,\pi_{\phi+\sigma}): \\
	& \pi_k \in \mathbb{R}_{> 0}, k=1, \dots, \phi+\sigma \big\} \subset \mathbb{R}^{m\times m}.
\end{split}
\end{equation}
For $ \mathds{S}_{\Sigma\oplus \Delta} $, the inequality (\ref{T2}) holding is equivalent to state that there exist positive vectors $ u, \nu, \xi,\zeta\in \mathbb{R}^N_{> 0} $ and a matrix $ \varPi\in \mathds{K} $, such that the following inequalities hold:
\begin{equation}\label{C2}
\left\{
\begin{split}
	&\sqrt{\varepsilon_2}\varPi^{1/2}F \xi <\nu,\\
	&(K(-I_N+\bar{\varGamma}H^{-1}G) + \varsigma_2 I_N)\xi \\
	& ~+ \sqrt{\varepsilon_2 }K \bar{\varGamma} H^{-1} E \varPi^{-1/2}u < 0,\\
	&\sqrt{\varepsilon_2}\varPi^{-1/2}E^\mathrm{T} H^{-1} \bar{\varGamma}K \zeta<u,\\
	&((-I_N + G^\mathrm{T} H^{-1} \bar{\varGamma} ) K + \varsigma_2 I_N)\zeta \\
	&~+ \sqrt{\varepsilon_2 }F^\mathrm{T} \varPi^{1/2}\nu < 0.
\end{split}
\right.
\end{equation}
\end{lemma}	

\textbf{Proof.} If Eq. (\ref{T2}) holds, the following system
\begin{align}
\mathds{S}^2_\varsigma:
\left\{
\begin{aligned} \nonumber
&\dot{x}=
\begin{aligned}[t]
&(K(-I_N+\bar{\varGamma}H^{-1}G) + \varsigma_2 I_N)x \\
& ~+ \sqrt{\varepsilon_2}K\bar{\varGamma} H^{-1}E u, \\
\end{aligned}\\
&y= 
\begin{aligned}[t]
&\sqrt{\varepsilon_2}F x,\\
\end{aligned}\\
\end{aligned}
\right.
\end{align}
with feedback (\ref{U1}) is internally stable, for all $ \Delta\in \bm\Delta $ satisfying $ \| \Delta \|_2\leq 1 $. Denote $ M^2_\varsigma(s) $ as the transfer function of $ \mathds{S}^2_\varsigma $. According to \cite[Theorem 10]{colombino2015convex}, there exists a matrix $ \varPi \in \mathds{K} $ such that $ \|\varPi^{1/2}{M}^2_\varsigma(0)\varPi^{-1/2}\|_2<1 $. When Lemma \ref{Y-lem-2} is applied, there exist vectors $u,v\in \mathbb{R}_{> 0}^N$ such that
\begin{equation}\label{LMI10}
  \begin{cases}
    -\varPi^{1/2} \sqrt{\varepsilon_2} F (K(-I_N+\bar{\varGamma}H^{-1}G) \\
    ~+ \varsigma_2 I_N)^{-1}\sqrt{\varepsilon_2} K \bar{\varGamma} H^{-1}E\varPi^{-1/2}u < \nu,\\
    -\varPi^{-1/2}\sqrt{\varepsilon_2} E^\mathrm{T} H^{-1} \bar{\varGamma} K ((-I_N + G^\mathrm{T} H^{-1} \bar{\varGamma} ) K \\
    ~+ \varsigma_2 I_N)^{-1} \sqrt{\varepsilon_2}F^{\mathrm{T}}\varPi^{1/2}\nu < u.
  \end{cases}
\end{equation}
The conditions in (\ref{C2}) can be obtained by applying Lemma \ref{Y-lem-3} to the inequalities in (\ref{LMI10}). The proof of sufficiency is obvious due to reversibility and thus is omitted.
$\hfill \hfill \blacksquare $
\vspace{0.2cm}

\begin{assumption}\label{Y-ass-3}
There exists $R(\theta)={\rm diag}(r_1(\theta),\dotsc,r_{n}(\theta))$ where $ r_i(\theta) $ are monomial diagonals, such that each element of $\tilde{A}(\theta)={A}(\theta)+R(\theta)$ is a posynomial of $\theta$ or zero. Moreover, each element of ${B}(\theta)$ and ${C}(\theta)$ is a posynomial of $\theta$ or zero.
\end{assumption}
\begin{assumption}\label{Y-ass-4}
There exists a constant $L_0$ such that $\tilde{L}(\theta)={L}(\theta)+L_0$ is a posynomial.
Moreover, there exist posynomials $\{f_i(\theta_i)\}_{i\in \emph{\textbf{I}}[1,p]}$ such that the constraint set $\Theta$ is equal to $\{\theta\in \mathbb{R}^{\theta}\mid \theta>0,~f_i(\theta_i)\leq 1, \forall i\in \emph{\textbf{I}}[1,p]\}$.
\end{assumption}

Then, the crucial result of the above analyses can be summarized in the following lemma:

    \begin{lemma}\label{Y-the-1}
Under Assumptions \ref{Y-ass-3} and \ref{Y-ass-4}, the robust stabilization problem (\ref{Problem10}) can be solved by the following GP in standard form:
	\begin{align*}
\mathds{P}_{2}:&~~
    \min_{\begin{subarray}{c}
    \theta\in \mathbb{R}_{> 0}^{n_{\theta}},\\
    \rho \in \mathbb{R}_{> 0}^N, \varPi \in \mathds{K} \\
    u, v, \xi, \zeta \in \mathbb{R}_{> 0}^{N} \end{subarray}} \quad \tilde{L}(\theta) \\
    {\rm s.t.}~
	&D^{-1}_\rho K^{-1}((K \bar{\varGamma} H^{-1} G + \varsigma_1 I_N)^\mathrm{T} \rho \\
	& ~+ \sqrt{\varepsilon_1}F^\mathrm{T} \boldsymbol{1}_N )<\boldsymbol{1}_N, \\
	&\sqrt{\varepsilon_1}E^\mathrm{T} H^{-1} \bar{\varGamma} K \rho<\boldsymbol{1}_N, \\
	&\sqrt{\varepsilon_2} D^{-1}_{\nu}\varPi^{1/2}F \xi <\boldsymbol{1}_N,\\
	&D_{\xi}^{-1} K^{-1}(K \bar{\varGamma} H^{-1} G\xi + \varsigma_2 \xi \\
	& ~+ \sqrt{\varepsilon_2 }K \bar{\varGamma} H^{-1} E \varPi^{-1/2}u )<\boldsymbol{1}_N,\\
&\sqrt{\varepsilon_2}D^{-1}_{u}\varPi^{-1/2}E^\mathrm{T} H^{-1} \bar{\varGamma} K\zeta<\boldsymbol{1}_N,\\
	&D^{-1}_{\zeta}K^{-1}(G^\mathrm{T} H^{-1} \bar{\varGamma}K\zeta + \varsigma_2 \zeta  \\
	& ~+ \sqrt{\varepsilon_2 }F^\mathrm{T} \varPi^{1/2}\nu) <\boldsymbol{1}_N, \\
    &~\frac{h_i-\underline{h}}{\bar{h}-\underline{h}}\leq\boldsymbol{1},
     ~\frac{\bar{h}-h_i}{\bar{h}-\underline{h}}\leq \boldsymbol{1},\ \forall i\in \textbf{I}[1,N],\\
    &~\frac{g_i-\underline{g}}{\bar{g}-\underline{g}}\leq \boldsymbol{1},
     ~\frac{\bar{g}-g_i}{\bar{g}-\underline{g}}\leq \boldsymbol{1},\ \forall i\in \textbf{I}[1,N],\\
    &f_i(\theta)\leq \boldsymbol{1}, ~~\forall i\in \emph{\textbf{I}}[1,p],
	\end{align*}
    \end{lemma}
where $ \tilde{L}(\theta) = L(\theta) + L_0 $.
			
    \textbf{Proof.} Lemmas \ref{Y-pro-2} and \ref{Y-pro-3} show that the robust stabilization problem $ \mathds{P}_{2} $ can be solved via the following optimization problem:
    \begin{align*}\label{Problem20}
    \mathds{P}_{10}:~~
    \min_{\begin{subarray}{c}
    \theta\in \mathbb{R}_{> 0}^{n_{\theta}},\\
    \rho \in \mathbb{R}_{> 0}^N, \varPi \in \mathds{K} \\
    u, v, \xi, \zeta \in \mathbb{R}_{> 0}^{N} \end{subarray}} \quad &L(\theta) \\
    \rm s.t.\quad~~~~ &(\ref{C1})~{\rm and}~ (\ref{C2}). \nonumber
    \end{align*}
By recalling Assumptions \ref{Y-ass-3} and \ref{Y-ass-4}, as well as the constraints of gains $ g_{i} $ and $ h_i $, the above problem can be converted into a standard GP optimization problem via simple algebraic manipulation as shown in Lemma \ref{Y-the-1}.
$\hfill \hfill \blacksquare $
\subsection{A GP Solution to the GoG}
After being maliciously attacked by an adding-edge attacker, the relevant linearized dynamics that follow from 
Eq. (\ref{FM2}) can be written as:
\begin{equation}\label{Y-FM4}
x(k+1) \!=\!
\left(K\left(-I_N \!+\! \bar{\varGamma} {\rm diag}(1/h)
\tilde{\boldsymbol{A}} {\rm diag}(g) \right) \!+\! I_N \right)x(k).
\end{equation}
Then, the condition to ensure the robust stability of system (\ref{Y-FM4}) is similar to that of system (\ref{Y-FM1}):
\begin{equation}\label{linearmodel}
{\rm sup}~ \lambda_{\max}\left(K\left(-I_N \!+\! \bar{\varGamma} {\rm diag}(1/h)  \tilde{\boldsymbol{A}} {\rm diag}(g) \right) \right)<0.
\end{equation}
By applying $\tilde{\boldsymbol{A}}=\boldsymbol{A}+\boldsymbol{A}_q$ and introducing a small positive scalar $ \varsigma\in (0, 1) $, we obtain the following condition:
\begin{align*}
  \sup_{
  \scriptsize \begin{aligned}
  &\boldsymbol{A}_q\in \mathcal{C}\setminus\mathcal{E},\\
  &\|\boldsymbol{A}_q\|_1\leq \bar{q}_1,\\
  &\|\boldsymbol{A}_q\|_2\leq \bar{q}_2
  \end{aligned}
  }
  \begin{aligned}
\lambda_{\max} ( K( -I_N \!+\! \bar{\varGamma} & {\rm diag}(1/h)   (\boldsymbol{A} \\
 & + \boldsymbol{A}_q ) {\rm diag}(g) ) ) \!<\! -\varsigma.
\end{aligned}
\end{align*}
Furthermore, the inequality constraints in (\ref{Cons1}) can easily be transformed into posynomial constraints.

There have two subnetwork policymakers $ P_{\pi}$, $ \pi = 1, 2 $, and the adding-edge attacker in our framework of the Stackelberg game. 
The ultimate goal of the two subnetwork policymakers is to minimize the global cost function at the required SINR. Their decision making can be described as a game-theoretical problem in a round-robin manner:
\begin{flalign}
  \mathds{P}_{\pi}^r: &\min_{\theta(r+c_\pi)}\ \
  \max_{ \scriptsize \begin{aligned}
  &~\boldsymbol{A}_q\in \mathcal{C}\setminus\mathcal{E},&\\
  &\|\boldsymbol{A}_q\|_1\leq \bar{q}_1,\\
  &\|\boldsymbol{A}_q\|_2\leq \bar{q}_2
  \end{aligned} }~~
  \sum_{i=1}^2{L}_i(\theta(r+c_\pi))\label{SGgamma}&\\
  {\rm s.t.}
  &~\lambda_{\max}\left(K\left(-I_N \!+\! \bar{\varGamma} {\rm diag}(1/h) \left(\boldsymbol{A} \!+\! \boldsymbol{A}_q\right) {\rm diag}(g) \right) \right) \nonumber&\\
  &~< -\varsigma,\tag{\ref{SGgamma}{a}} \label{SGgammaa}&\\
  &~\frac{h_i-\underline{h}}{\bar{h}-\underline{h}}\leq 1,~\frac{\bar{h}-h_i}{\bar{h}-\underline{h}}\leq 1,\ \forall i\in \textbf{I}[1,N],\tag{\ref{SGgamma}{b}} \label{SGgammab}&\\
  &~\frac{g_i-\underline{g}}{\bar{g}-\underline{g}}\leq 1,~\frac{\bar{g}-g_i}{\bar{g}-\underline{g}}\leq 1,\ \forall i\in \textbf{I}[1,N],\tag{\ref{SGgamma}{c}} \label{SGgammac}&\\
  &~\theta_j(r+c_\pi)=\theta_j(r), ~~\forall j\in \mathcal{V}_{-\pi}, \pi = 1, 2, \tag{\ref{SGgamma}{d}} \label{SGgammad}&
 \end{flalign}
where $c_\pi\in \mathbb{Z}_{>0}$ denotes the policy update frequency of the $\pi$th subnetwork policymaker.

\begin{remark}
Condition (\ref{SGgammad}) shows the lack of coordination between two policymakers in the Nash game because of the privacy considerations for cellular users in the two subnetworks; both make decisions without anticipating the real-time strategy of the other. To prevent a potential endless loop, it is recommended to employ a series of $c_\pi$ primes to each other.$\hfill \hfill \square $
\end{remark}

For the attacker who increases the interference among cellular users by changing the network topology, she aims to maximize the cost function that the policymakers use to improve the QoS, the strategy can be described as
\begin{equation*}\label{ProblemA}
\mathds{P}_{A}: \max_{ \scriptsize \begin{aligned}
  &~\boldsymbol{A}_q\in \mathcal{C}\setminus\mathcal{E},\\
  &\|\boldsymbol{A}_q\|_1\leq \bar{q}_1,\\
  &\|\boldsymbol{A}_q\|_2\leq \bar{q}_2
  \end{aligned}
  }\sum_{i=1}^2{L}_i(\theta^*),
\end{equation*}
where $\theta^*$ represents the optimal policy of two policymakers obtained via $ \mathds{P}_{\pi}^r $.

Denote Eq. (\ref{SGgamma}) as a ternary set $\mathds{D}_{\pi}=\{\mathcal{P}, ({\Theta}_{\pi}, \mathds{G}), L_{\rm{sum}}\}$, consisting of the player set $\mathcal{P}=\{P_{1},P_{2},P_A\}$, the action spaces $({\Theta}_{\pi},\mathds{G})$, and the objective function $L_{\rm{sum}}=L_1(\theta_1)+L_2(\theta_2)$. The interaction between the two subnetworks under the threat of the attacker can be captured through a Nash game, which is denoted as
$\mathds{D}_{N}=\{(P_1,P_2), ({\Theta}_{1}, {\Theta}_{2}), L_{\rm{sum}}\}$. Both $ P_1 $ and $ P_2 $ are intended to minimize the total cost $ L_{\rm{sum}} $.
The Nash game $\mathds{D}_{N}$ together with the Stackelberg game $\mathds{D}_{\pi}$ constitute a Gestalt game \cite{chen2018security} shown in Fig. \ref{fig:figure2}.

We study the robust stabilization and resource allocation problems of the FM algorithm under various kinds of structured uncertainties. The existence of the attacker can be considered as a worst-case scenario in uncertainties.
Thus, the results in Lemma \ref{Y-the-1} can be extended to the resilient containment problem of interference in communication networks under the threat of adding-edge attacks.
Next, we introduce the main theorem of the current work:

\begin{myTheo}\label{Y-the-2}
The optimal policy seeking problem $\mathds{P}_{\pi}^r$ is equivalent to the following round-robin GP problem:
\begin{align}
\mathds{Q}_{\pi}^r: &\min_
{\scriptsize
\begin{aligned}
\theta_\pi(r+c_\pi&)\in \mathbb{R}_{> 0}^{2N_\pi},\\
\rho \in \mathbb{R}_{> 0}^N&, \varPi \in \mathds{K} \\
u, v, \xi, \zeta &\in \mathbb{R}_{> 0}^{N}
\end{aligned}}
\ \ \max_{ \scriptsize \begin{aligned}
  \boldsymbol{A}_q\in \mathcal{C}\setminus&\mathcal{E},\\
  \|\boldsymbol{A}_q\|_1\leq &\bar{q}_1,\\
  \|\boldsymbol{A}_q\|_2\leq& \bar{q}_2
  \end{aligned}
  }~~\sum_{i=1}^2\tilde{L}_i(\theta(r+c_\pi))\label{GPgamma}\\
  {\rm s.t.}
  &~D^{-1}_\rho K^{-1} [( K \bar{\varGamma} {\rm diag}(1/h) \boldsymbol{A}  {\rm diag}(g) + \varsigma I_N )^\mathrm{T} \rho \nonumber\\
  & ~+ \sqrt{\bar{q}_1}  {\rm diag}(g) \boldsymbol{1}_N ]<\boldsymbol{1}_N, \tag{\ref{GPgamma}{a}} \label{GPgammaa}\\
  &~\sqrt{\bar{q}_1} {\rm diag}(1/h) \bar{\varGamma} K \rho <\boldsymbol{1}_N,  \tag{\ref{GPgamma}{b}} \label{GPgammab}\\
  &~\sqrt{\bar{q}_2} D^{-1}_{\nu}\varPi^{1/2} {\rm diag}(g) \xi <\boldsymbol{1}_N,  \tag{\ref{GPgamma}{c}} \label{GPgammac}\\
  &~D_{\xi}^{-1} K^{-1} ( K \bar{\varGamma} {\rm diag}(1/h) \boldsymbol{A} {\rm diag}(g) \xi + \varsigma\xi \nonumber\\
  & ~+ \sqrt{\bar{q}_2 }K \bar{\varGamma}
  {\rm diag}(1/h) \varPi^{-1/2}u )<\boldsymbol{1}_N,
  \tag{\ref{GPgamma}{d}} \label{GPgammad}\\
  &~\sqrt{\bar{q}_2}D^{-1}_{u}\varPi^{-1/2} {\rm diag}(1/h) \bar{\varGamma} K\zeta<\boldsymbol{1}_N,
  \tag{\ref{GPgamma}{e}} \label{GPgammae}\\
  &~D^{-1}_{\zeta}K^{-1} ( {\rm diag}(g) \boldsymbol{A}^\mathrm{T} {\rm diag}(1/h) \bar{\varGamma}K\zeta + \varsigma\zeta \nonumber \\
  & ~+ \sqrt{\bar{q}_2 } {\rm diag}(g) \varPi^{1/2}\nu ) <\boldsymbol{1}_N,  \tag{\ref{GPgamma}{f}} \label{GPgammaf}\\
  &~\frac{h_i-\underline{h}}{\bar{h}-\underline{h}}\leq \boldsymbol{1},
   ~\frac{\bar{h}-h_i}{\bar{h}-\underline{h}}\leq \boldsymbol{1},\ \forall i\in \textbf{I}[1,N],
  \tag{\ref{GPgamma}{g}} \label{GPgammag}\\
  &~\frac{g_i-\underline{g}}{\bar{g}-\underline{g}}\leq \boldsymbol{1},
   ~\frac{\bar{g}-g_i}{\bar{g}-\underline{g}}\leq \boldsymbol{1},\ \forall i\in \textbf{I}[1,N],
  \tag{\ref{GPgamma}{h}} \label{GPgammah}\\
  &~\theta_j(r+c_\pi)=\theta_j(r), ~~\forall j\in \mathcal{V}_{-\pi}, \pi = 1, 2,
  \tag{\ref{GPgamma}{i}} \label{GPgammai}
  \end{align}
where $N_\pi$ is the node number of Network $\pi$, $\tilde{L}_i={L}_i+L_0$ with $L_0=N(\frac{\bar{g}^{-p}}{\underline{g}^{-p}-\bar{g}^{-p}}+
\frac{\underline{h}^{q}}{\bar{h}^{q}-\underline{h}^{q}})$.
\end{myTheo}

\textbf{Proof.}
Let
$ A_1 = K \left(-I_N+\bar{\varGamma}{\rm diag}(1/h) \boldsymbol{A} {\rm diag}(g) \right) $,
$ B_1 = K\bar{\varGamma} {\rm diag}(1/h) $,
$ C_1 = {\rm diag}(g) $, $ B_2 = \boldsymbol{A}_q $, $ C_2 = I_N$ and
$ A_2 = D_1 = D_2 = \boldsymbol{0}_n $. By recalling Lemma \ref{Y-the-1}, Condition (\ref{SGgammaa}) in $\mathds{P}_{\pi}^r$ with constraints $\|\boldsymbol{A}_q\|_1\leq \bar{q}_1$ and $\|\boldsymbol{A}_q\|_2\leq \bar{q}_2$ can easily be reduced to the Conditions (\ref{GPgammaa})
$ \sim $ (\ref{GPgammaf}) in $\mathds{Q}_{\pi}^r$.
Conditions (\ref{GPgammag})$ \sim $ (\ref{GPgammai}) in $\mathds{Q}_{\pi}^r$ and Conditions (\ref{SGgammab})$ \sim $ (\ref{SGgammad}) in $\mathds{P}_{\pi}^r$ are identical. It is proved that the total cost function $\tilde{L}(\theta)$ satisfies Assumption \ref{Y-ass-4} with  $L_0=N(\frac{\bar{g}^{-p}}{\underline{g}^{-p}-\bar{g}^{-p}}+
\frac{\underline{h}^{q}}{\bar{h}^{q}-\underline{h}^{q}})$.
$\hfill \hfill \blacksquare $
\vspace{0.2cm}

Due to insufficient coordination between the two subnetwork policymakers, the attack magnitude $ \bar{q}_2 $ that the whole topology can withstand cannot be large. Related to $ \bar{q}_2 $, we propose Theorem \ref{Y-the-3} to provide a sufficient condition such that the GP problem $\mathds{Q}_{\pi}^r$ can be solved.

\begin{myTheo}\label{Y-the-3}
  For a given $\bar{q}_1$, the game $\mathds{Q}_{\pi}^r$ is solvable for all $\bar{q}_2\in [0, \bar{q}_{2}^{*}]$, where $\bar{q}_{2}^{*}=\min\{\bar{q}_{2,\pi}\}$, with $\bar{q}_{2,\pi}$ given by the following GP problem:
  \begin{align}
    \mathds{R}_{\pi}: &\min_
    {\scriptsize
\begin{aligned}
\theta_\pi\in& \mathbb{R}_{> 0}^{2N_\pi},\\
\rho \in \mathbb{R}_{> 0}^{N},&~\Pi\in \mathds{K}, \\
u,v \in \mathbb{R}_{> 0}^{N},&~\xi,\zeta\in \mathbb{R}_{> 0}^{N}
\end{aligned}}
\ \ \max_{\scriptsize
    \begin{aligned}
      \boldsymbol{A}_q & \in \mathcal{C}\setminus\mathcal{E},\\
     \bar{q}_{2,\pi}&\in \mathbb{R}_{> 0}
    \end{aligned}
    }
~~\frac{1}{\bar{q}_{2,\pi}}\nonumber\\
{\rm s.t.}
&~~~(\ref{GPgammaa}) \thicksim (\ref{GPgammah}),\nonumber\\
&~~~~\theta_j= \theta_j(0),\ \forall j\in \mathcal{V}_{-\pi}, \pi = 1, 2.\label{inititr}
\end{align}
\end{myTheo}

\textbf{Proof.}
Define $f_{\lambda}(q_2) = \lambda_{\max} \left( K \left(\! -I_N+\bar{\varGamma}{\rm diag}(1/h) \left( \boldsymbol{A} \! +\! \boldsymbol{A}_q(q_2) \right) {\rm diag}(g) \right) \right)$
with $q_2\leq \bar{q}_2$. For a fixed $\bar{q}_1$, $ f_{\lambda}(q_2) $ is a monotonically increasing function.
If $\bar{q}_2$ is excessively large, it will be extremely hard to satisfy Conditions (\ref{GPgammac})$\sim$(\ref{GPgammaf}), so $\mathds{Q}_{\pi}^r$ cannot be solved.
The worst case of the $\pi$th subnetwork means that the maximization of $\bar{q}_2$ is achieved only by the $\pi$th policymaker without any assistance from the other policymaker. Condition (\ref{inititr}) can be used to solve $ \bar{q}_{2,\pi}$, $\pi = 1, 2 $, which indicates the noncooperation between the two subnetworks. Hence, one can obtain that the maximum attack magnitude is $\bar{q}_{2}^{*}=\min_{\pi}\{\bar{q}_{2,\pi}\}$.
$\hfill \hfill \blacksquare $
\vspace{0.2cm}

In the following, as a direct corollary of Theorem \ref{Y-the-3}, we propose a property of the attacker such that the problem $\mathds{Q}_{\pi}^r$ is always feasible.

\begin{myCollo}\label{Theo50}
For a given $\bar{q}_1$, games $\mathds{Q}_{1}^r$ and $\mathds{Q}_{2}^r$ are solvable for the attack magnitude $\bar{q}_2\in (0, \bar{q}_{2}^{*}]$, where $\bar{q}_{2}^{*}$ is obtained by solving problem $\mathds{R}_{\pi}$.
\end{myCollo}%

\textbf{Proof.}
Since the process of solving $\mathds{Q}_{1}^r$ and $\mathds{Q}_{2}^r$ is the same, we take $\mathds{Q}_{1}^r$ as an example for illustration. The first iteration $\mathds{Q}_{1}^0$ can be solved by recalling Theorem \ref{Y-the-3}, so that it can be ensured that each round of $\mathds{Q}_{1}^r$, $\forall r\in \mathbb{Z}_{>0}$, is solvable since $ P_1 $ adopts a continuous optimization solution. Thus, for a given topology satisfying the attack range $\bar{q}_2\in (0, \bar{q}_{2}^{*}]$, there must exist a feasible solution for $\mathds{Q}_{1}^r$.
$\hfill \hfill \blacksquare $
\vspace{0.2cm}

Based on the above discussions, the GP formulation of the FM algorithm with two subnetworks has been clearly expounded. Next, we focus on how to determine the GNE of the GoG in this work.
In the process of allocating limited engineering resources, both policymakers $P_\pi$, $ \pi=1, 2 $, obtain the allocation policy of all nodes on Network $ \pi $ by solving $\mathds{Q}_{\pi}^r$.
It should be noted that the current action $\theta_{\pi}^*$ of the $ \pi $th subnetwork policymaker is the best response associated with other policymakers at a certain previous updating steps and the worst-case attack strategy $ \boldsymbol{A}^*_q\in \mathcal{C}\setminus\mathcal{E} $.
Specifically, $P_\pi$ reassigns resources to all other nodes on Network $ \pi $ in every $c_\pi$ time interval.
Both policymakers $P_\pi$ focus on minimizing the unified total cost function by anticipating the worst-case attack at each time step in a non-fully cooperative manner.
One of the most direct methods to solve the GNE is to iteratively address $\mathds{Q}_{\pi}^r$ until $P_\pi$ chooses to follow the present policy.
Hence, we propose a round-robin algorithm to compute the exact GNE, whose contents are described in Algorithm \textbf{HIG}\footnotemark[1]\footnotetext[1]{One suitable strategy for Step 15 is shown in Algorithm $\mathbf{HWA}$.}. Notably, Algorithm \textbf{HIG} is asymptotically convergent according to the above analyses.

\begin{algorithm}[!]
  \floatname{algorithm}{Algorithm HIG}
  \caption*{$\mathbf{Algorithm}$ $\mathbf{HIG}$:~~Heuristic Iterative Algorithm for $\mathbf{Problem}$ $\mathbf{GSA}$} 
  {\bf Input:} 
  Stochastic nominal network topology ${\boldsymbol{A}}$. The threshold values of the parameters $\bar{g}$, $\underline{g}$, $\bar{h}$ and $\underline{h}$. 
  The attack magnitude $(\bar{q}_1,\bar{q}_2)$. All the updating frequencies $c_\pi$, $\pi=1,2$.\\
  {\bf Output:} 
  The data of the Gestalt Nash equilibrium, containing the  best response $\theta_{\pi}^*$ of $ P_\pi $, $\pi=1,2$, and the corresponding worst-case attack strategy $\boldsymbol{A}^*_q$.
  \begin{algorithmic}[1]
  \State Initialize action $\theta_{\pi}(0)$, the round counter $r=1$, iteration tolerance $t=10^{-6}$.
  \While {$r\leq \max\{c_1,c_2\}\vee \|\theta_1(r)-\theta_1(r-c_1)\|>t \vee \|\theta_2(r)-\theta_2(r-c_2)\|>t$}
  \If {$r\ {\rm mod}\ c_1=0$}
      \State $P_1$ obtains new strategy $\theta_{1}$ by solving $\mathds{Q}_{1}^r$.
  \EndIf
  \State \hspace*{-0.05in} \textbf{endif}
  \If {$r\ {\rm mod}\ c_2=0$}
      \State $P_2$ obtains new strategy $\theta_{2}$ by solving $\mathds{Q}_{2}^r$.
  \EndIf
  \State \hspace*{-0.05in} \textbf{endif}
  \State $r=r+1$.
  \EndWhile
  \State \hspace*{-0.05in} \textbf{endwhile}
  \State For $\theta^*_{\pi}$, obtain the associated worst-case attack policy $\boldsymbol{A}^*_q$ via heuristic strategies.
  \State \Return $\theta^*_{\pi}$, $\pi=1,2$, and $\boldsymbol{A}^*_q$.
  \end{algorithmic}
  \end{algorithm}

\begin{myTheo}\label{Theo100}
For a given topology that satisfies the attack magnitude $\bar{q}_2\in (0, \bar{q}_{2}^{*}]$, Algorithm \textbf{HIG} asymptotically converges to a GNE.
\end{myTheo}

\textbf{Proof.}
During a given iteration, the value of the current total cost function $\tilde{L}_{\rm{sum}}(r+c_\pi)=\sum_{i=1}^2\tilde{L}_i(\theta_i(r+c_\pi))$ does not exceed the value in the previous updating step $\tilde{L}_{\rm{sum}}(r)$, and the resulting nonincreasing sequence $\{\tilde{L}_{\rm{sum}}(r)\}$ is lower bounded by $L_0$.
According to the monotone convergence theorem \cite{ganter2012formal}, we can obtain the asymptotic convergence of $\{\tilde{L}_{\rm{sum}}(r)\}$.
Let the policy profile $ (\theta^*_{1},\theta^*_{2}) $ denote the optimal response under the worst-case attack strategy $\boldsymbol{A}^*_q$, which means that $ P_\pi $, $ \pi = 1,2 $, together with realizing the minimum of $ \tilde{L}_{\rm{sum}} $ at this point, that is,  $L(\theta^*_{1},\theta^*_{2})\leq L(\theta^*_{1},\theta_{2})$, $\forall \theta_{2}\in \Theta_2$, and $L(\theta^*_{1},\theta^*_{2})\leq L(\theta_{1},\theta^*_{2})$, $\forall \theta_{1}\in \Theta_1$.
We know that $(\theta^*_{1},\theta^*_{2})$ is a Nash equilibrium according to Definition \ref{Y-def-1}; together with $\boldsymbol{A}_q^*$, Algorithm \textbf{HIG} converges to a GNE. Moreover, $\boldsymbol{A}_q^*$ can be obtained from Algorithm \textbf{HWA} proposed later.
$\hfill \hfill \blacksquare $
\vspace{0.2cm}

\subsection{Adversarial Analysis}
From the perspective of the attacker, she adopts a relatively conservative attack method for better covertness, so the attack magnitude should not be too large.
Simultaneously, the attacker is often restricted by geographic location, equipment performance and other conditions, so that the attack range is limited. Therefore, we propose a reasonable conjecture about the means of attacks to obtain the attack strategy as soon as possible.

\begin{assumption}\label{assumption20}
The attacker only adds edges originating from one node.
\end{assumption}

\begin{remark}\label{Y-rmk5}
The node in Assumption \ref{assumption20} is recorded as the attack source and is denoted by $\mathcal{N}_a$, such as node $9$ in Fig. \ref{fig:figure1}. This indicates that the attacker selects a simple and efficient method to compromise the topology, that is, adding edges from a common node simultaneously.$\hfill \hfill \square $
\end{remark}

Under Assumption \ref{assumption20}, one can obtain that $\|\boldsymbol{A}_q\|_{1}=\frac{1}{2}\|\boldsymbol{A}_q\|_{L_1}$, where $\boldsymbol{A}_q \subset \mathcal{C}\setminus \mathcal{E}$ is the alternating adjacency matrix.
Both sides of the equation represent the sum of the weights of the added edges; in addition, the added amount can be conveniently counted via the $ L_1 $-norm.
Let $z\in \mathbb{Z}_{>0}$ denote the exact number of added edges.
We have $\boldsymbol{A}_q=\sum_{m=1}^{z}\textbf{e}_i^m({\textbf{e}}_{i,j}^m)^{\mathrm{T}}
+\textbf{e}_j^m({\textbf{e}}_{j,i}^m)^{\mathrm{T}}$, 
where $\textbf{e}_i^m$ and $\textbf{e}_{i,j}^m$ corresponding to the $m$th added edge respectively represent the N-dimensional zero vector with the $i$th entry being $1$ and the N-dimensional vector with edge weight $a_{ij}^m$.
Denote $\bar{\mu}$ as a right Perron vector of $K \left(-I_N+\bar{\varGamma}{\rm diag}(1/h) \boldsymbol{A} {\rm diag}(g) \right)$, one can obtain that
\begin{align*}
  \bar{\mu}^{\mathrm{T}}K & \left(-I_N+\bar{\varGamma}{\rm diag}(1/h) \boldsymbol{A} {\rm diag}(g)\right)\bar{\mu}\\
  & ~=\lambda_{\max}\left(K\left(-I_N+\bar{\varGamma}{\rm diag}(1/h) \boldsymbol{A} {\rm diag}(g) \right)\right).
\end{align*}
Furthermore, we obtain
\begin{align*}
\lambda_{\max}&\left(K\left(-I_N+\bar{\varGamma}{\rm diag}(1/h) \tilde{\boldsymbol{A}} {\rm diag}(g) \right)\right)\nonumber\\
\geq&~ \bar{p}^{\mathrm{T}}K\left(-I_N+\bar{\varGamma}{\rm diag}(1/h) \tilde{\boldsymbol{A}} {\rm diag}(g) \right)\bar{\mu} \nonumber\\
=&~\bar{\mu}^{\mathrm{T}}\left(K\left(-I_N+\bar{\varGamma}{\rm diag}(1/h) \boldsymbol{A} {\rm diag}(g) \right)\right)\bar{\mu}\\
&+\bar{\mu}^{\mathrm{T}}\left(K\bar{\varGamma}{\rm diag}(1/h) {\boldsymbol{A}}_q {\rm diag}(g) \right)\bar{\mu}\nonumber\\
=&~\lambda_{\max}\left(K\left(-I_N+\bar{\varGamma}{\rm diag}(1/h) \boldsymbol{A} {\rm diag}(g) \right)\right)\nonumber\\
&+\bar{\mu}^{\mathrm{T}}(K\bar{\varGamma}{\rm diag}(1/h) ( \sum_{m=1}^{z}\textbf{e}_i^m({\textbf{e}}_{i,j}^m)^{\mathrm{T}}\\
&+\textbf{e}_j^m ({\textbf{e}}_{j,i}^m)^{\mathrm{T}} ) {\rm diag}(g) ) \bar{\mu} \nonumber\\
=&~\lambda_{\max}\left(K\left(-I_N+\bar{\varGamma}{\rm diag}(1/h) \boldsymbol{A} {\rm diag}(g) \right)\right)\\
&+\sum_{m=1}^{z}a_{ij} \left(k_i \bar{\gamma}_i g_i / h_i + k_j \bar{\gamma}_j g_j / h_j\right)\bar{\mu}_i\bar{\mu}_j,\nonumber
\end{align*}
where $\bar{\mu}_i$ denotes the $i$th element of $\bar{\mu}$. To maximize network connectivity to enhance the impact of interference, the attacker tends to add edges with larger $\left(k_i \bar{\gamma}_i g_i / h_i + k_j \bar{\gamma}_j g_j / h_j\right) \bar{\mu}_i\bar{\mu}_{j}$ based on the above formula. Thus, the added edge $E^*(i,j)$ of the best attack to compromise the whole topology satisfies
\begin{equation}\label{attstra}
E^*(i,j)\in {\rm arg} \max_{E(i, j)\in \mathcal{C}\setminus \mathcal{E}} \left(\frac{k_i \bar{\gamma}_i g_i}{h_i} + \frac{k_j \bar{\gamma}_j g_j}{h_j}\right) \bar{\mu}_i\bar{\mu}_j.
\end{equation}
The above worst-case attack-seeking strategy can be regarded as the greedy heuristic algorithm described in Algorithm \textbf{HWA}. Specifically, the attacker can add a set of key edges by iteratively employing the strategy (\ref{attstra}) that affects the topology to the greatest extent.

\begin{remark}\label{Y-rmk3}
Algorithm \textbf{HIG} implies that two subnetwork policymakers can obtain their optimal strategy $\theta^*_{\pi}$ without anticipating the exact edges added by the malicious attacker. However, it can be seen from Algorithm \textbf{HWA} that the attacker can only obtain the optimal strategy $\boldsymbol{A}_q^*$ based on the policy set of all network policymakers. These factors reflect the difference between two subnetwork policymakers and the attacker in a general SG framework.$\hfill \hfill \square $
\end{remark}

\begin{algorithm}[!]
  \floatname{algorithm}{Algorithm HWA}
  \caption*{$\mathbf{Algorithm}$ $\mathbf{HWA}$:~~Heuristic Algorithm for the Worst-case Attacker} 
  {\bf Input:} 
  Stochastic nominal network topology ${\boldsymbol{A}}$. The Nash equilibrium of the two policymakers $(\theta_{1}^*, \theta_{2}^*)$. The attack magnitude $(\bar{q}_1,\bar{q}_2)$. \\
  {\bf Output:} 
  The corresponding worst-case attack strategy $\boldsymbol{A}^*_q$.
  \begin{algorithmic}[1]
  \State Initialize $\boldsymbol{A}^*_q=O_{N\times N}$, $\tilde{\boldsymbol{A}}=\boldsymbol{A}$, and the set of attack sources
  $\mathcal{N}_a=\mathcal{V}$.
  \While {$\|\boldsymbol{A}^*_q\|_{2}< \bar{q}_2\wedge\|\boldsymbol{A}^*_q\|_{1}< \bar{q}_1$}
  \State Choose the best edge $E^*(i,j)$ according to (\ref{attstra}), where $\bar{p}_i$ is computed from $K \left(-I_N+\bar{\varGamma} {\rm diag}(1/h^*) \tilde{\boldsymbol{A}} {\rm diag}(g^*) \right)$ and either node $i$ or $j$ belongs to $\mathcal{N}_a$.
  \State Update $\boldsymbol{A}^*_q$ by adding the edge $E^*(i,j)$ weighted by $1$.
  \State Update $\tilde{\boldsymbol{A}}=\boldsymbol{A}+\boldsymbol{A}^*_q$.
  \State Update $\mathcal{N}_a=\mathcal{N}_a\cap\{i,j\}$.
  \State Update the last added edge ${E}_f=E^*(i,j)$.
  \EndWhile
  \State \hspace*{-0.05in} \textbf{endwhile}
    \If{$\|\boldsymbol{A}^*_q\|_{1}\geq \bar{q}_1$}
  \State Update $\boldsymbol{A}^*_q$ by changing the weight of ${E}_f$ to $w_f\leq1$ such that $\|\boldsymbol{A}^*_q\|_1= \bar{q}_1$.
  \EndIf
  \State \hspace*{-0.05in} \textbf{endif}
  \If{$\|\boldsymbol{A}^*_q\|_{2}\geq \bar{q}_2$}
  \State Update $\boldsymbol{A}^*_q$ by changing the weight of ${E}_f$ to $w_f\leq1$ such that $\|\boldsymbol{A}^*_q\|_2= \bar{q}_2$.
  \EndIf
  \State \hspace*{-0.05in} \textbf{endif}
  \State Update $\tilde{\boldsymbol{A}}=\boldsymbol{A}+\boldsymbol{A}^*_q$.
  \State \Return $\boldsymbol{A}^*_q$.
  \end{algorithmic}
  \end{algorithm}

\section{Numerical Simulations}	\label{section4}

In this section, we illustrate the feasibility of the above theoretical results via several numerical simulations. 
As shown in Fig. \ref{fig:figure3}, we consider a network topology $ \mathcal{G}$ ($N=22$) consisting of two subnetworks, that is, Networks $ 1 $ and $ 2 $. Specifically, Network $1$ consists of user equipment indexed by $1\sim 11$, while Network $2$ consists of user equipment indexed by $12\sim 22$. The blue solid lines and blue dotted lines represent inter-edges and intra-edges between the two subnetworks, respectively. 
All blue edges on the nominal graph $\mathcal{G}$ have a weight of $1.0$. Moreover, the edges added by the attacker are depicted by red dashed lines. 

Similar to the setting in \cite{preciado2014optimal}, the interference gain from channel $ j $ to channel $i$ is adjusted to $ g_{ji} $ by adopting the following cost function:
\begin{equation}\label{Y-alphag}
  \alpha(g_{i})=\frac{g_{i}^{-p}-\bar{g}^{-p}}{\underline{g}^{-p}-\bar{g}^{-p}},
\end{equation}
where $ p \in \mathbb{R}_{>0} $. Note that $ \alpha(g_{i}) $ is normalized as $\alpha(\underline{g})=1$ and $\alpha(\bar{g})=0$. Likewise, the following cost function is used to adjust the transmission gain to $ h_i $:
\begin{equation}\label{Y-betah}
  \beta(h_{i})=\frac{h_{i}^{q}-\underline{h}^{q}}{\bar{h}^{q}-\underline{h}^{q}},
\end{equation}
where $ q \in \mathbb{R}_{>0} $. Apparently, $\beta(\underline{h})=0$ and $\beta(\bar{h})=1$. The exact values of $p$ and $q$ can be obtained by data fitting on the experimental data set.

\subsection{Tolerable Attack Magnitude}

Select $\underline{g}=0.1$, $\bar{g}=0.9$, $\underline{h}=4$, and $\bar{h}=6$ in the cost functions (\ref{Y-alphag}) and (\ref{Y-betah}) with exponents $p=0.1$ and $q=1$. The curves of the cost functions $\alpha(g_i)$ and $\beta(h_i)$ are depicted in Fig. \ref{fig:figure5:a} and Fig. \ref{fig:figure5:b}, respectively. Set the parameter $ k_i = 1 $ in the FM algorithm, that is, $ K = I_N $.
The predefined convergent rate is selected as $\varsigma=0.01$. For simplicity, we assume that all channels have the same SINR and select $ \bar{\gamma_{i}} =1$.
The initial policy values of the two policymakers are listed in Table \ref{table1}.

For $\bar{q}_1=2.25$, the allowable maximum attack magnitude is computed as $\bar{q}_{2}^{*}=\min\{5.658$, $6.189\}=1.582\cdot\|\boldsymbol{A}\|$ by solving the GP in Theorem \ref{Y-the-3}.

\subsection{GNE Seeking}
Set $t=10^{-4}$, $c_1=2$, and $c_2=3$ in Algorithm \textbf{HIG}.
For various values of the sequence $\bar{q}_2\in [0, \bar{q}_{2}^{*}]$, we use GP in Theorem \ref{Y-the-2} to solve the non-fully cooperative resource distribution game and determine the optimal policies of both subnetwork policymakers in a Nash equilibrium sense.
As shown in Fig. \ref{fig:figure10}, we obtain the corresponding cost curve versus $\bar{q}_2 / \|\boldsymbol{A}\|$: 
\begin{enumerate}
\item When $\bar{q}_2 / \|\boldsymbol{A}\|$ is small, Network $1$ can achieve the required SINR with little investment. Thus, only Network $2$ increases its investment gradually to guarantee the QoS. Moreover, the cost of Network $1$ increases gradually as $\bar{q}_2 / \|\boldsymbol{A}\|$ increases.
\item As $\bar{q}_2 / \|\boldsymbol{A}\|$ increases, Network $1$ starts to invest to maintain the SINR and its cost rises almost linearly. Correspondingly, the cost increasing rate of the investment of Network $2$ slows down gradually.
\item When $\bar{q}_2 / \|\boldsymbol{A}\|$ is sufficiently large, Network $2$ is inclined to stop increasing its investment. In contrast, Network $1$ continues to increase its investment until the Nash equilibrium between the two subnetworks no longer holds.
\end{enumerate}

In Fig. \ref{fig:figure15}, we record the distribution of optimal investments versus the PageRank \cite{Berkhin2005A} of all nodes when the attack magnitude $\bar{q}_2=0.292\cdot\|\boldsymbol{A}\|$, $0.929\cdot\|\boldsymbol{A}\|$, $1.176\cdot\|\boldsymbol{A}\|$, and $1.492\cdot\|\boldsymbol{A}\|$. Generally speaking, the higher the PageRank of the node (user equipment), the greater the investment.
Almost all of the nodes receive greater investment as the attack magnitude $\bar{q}_{2}$ increases.
Moreover, the border nodes, such as nodes $15$ and $21$, receive greater investments than other nodes to enhance transmission in Cases $1$, $2$, and $3$, implying that the intra-edges between the two subnetworks are more vulnerable than the inter-edges generally. 

As shown in Fig. \ref{fig:figure15:a}, most user equipments do not receive any investment from policymakers when under an attack of low magnitude. The nodes with PageRank less than $0.035$ receive no investment.
In contrast, almost all of the nodes in Fig. \ref{fig:figure15:d} receive more than $0.300$ unit investment when subject to the maximum allowable attack magnitude. 
These simulation results show that identifying the specific connection situation of the nodes in the actual topology is of great significance for allocating communication resources among the nodes.
There is an interesting phenomenon in which the covariance between the adjacent $\theta^*$ values gradually decreases as $\bar{q}_2$ increases. One possible reason for this is that the policymaker cannot easily achieve its goal by investing in a small group of critical nodes on $ \mathcal{G} $ if the attacker is too aggressive.

Finally, we investigate how the compromised attack strategy of the attacker is effectively implemented. The edges added by the attacker, as well as the corresponding weights, are computed according to Algorithm \textbf{HWA} with the results shown in Table \ref{table2} (see the red dashed lines in Fig. \ref{fig:figure3} for better illustration to Case $4$).
It can be found that the attacker prefers to add the inter-edges from the nodes with high connectivity on Network $2$, such as nodes $13$, $15$, and $17$. One of the possible reasons is that the overall connectivity of Network $2$ is higher than that of Network $1$, and the interference caused by the added inter-edges to Network $2$ is more influential. For Case $4$, another possible reason is that the defense strategy of Network $1$ is strong and the investment is huge. The above strategies show that the attacker chooses the edges to add based on not only the whole nominal topology but also on the policy profiles of the two subnetwork policymakers, corroborating the analysis in Remark \ref{Y-rmk3}.

\section{Conclusion}\label{section5}


The problem of seeking optimal communication performance strategies on cellular networks under unknown malicious adding-edge attacks in the GoG framework is studied. The conflict between each subnetwork policymaker and the attacker is modeled as a Stackelberg game, while the lack of coordination between the two subnetwork policymakers is established by the means of a Nash game. We first demonstrate that the communication resource allocation of cellular networks based on the FM algorithm can be efficiently handled by GP. Then, combined with the attacking manner of the attacker, the tolerable maximum attack magnitude is investigated. A GNE is proven to exist within this attack range. Based on the robustness of the results obtained by the above GP, we propose a heuristic round-robin algorithm to compute the optimal policy profile for each subnetwork policymaker. Correspondingly, a greedy heuristic strategy for the attacker to disrupt the communication topology is also developed. The optimal policy profile of the two policymakers and the adding-edge strategy of the attacker together constitute the aforementioned GNE. Simulation examples illustrate that the proposed theoretical results can realize good robustness and successfully obtain the GNE. Future work will attempt to exploit reinforcement learning or automatic data processing to solve the GNE in this paper and extend the findings to more complex communication networks.

\bibliography{PIDFR}
\end{document}